\RequirePackage{fix-cm}
\documentclass[twocolumn]{svjour3}          
\smartqed  
\usepackage{graphicx}
\usepackage{mathptmx}
\usepackage{cite}
\usepackage{hyperref}
\usepackage{filecontents,lipsum}
\usepackage{subcaption}
\usepackage[skip=2pt,font=scriptsize]{caption}
\usepackage[skip=2pt,font=scriptsize]{subcaption}
\usepackage{amsmath,amssymb,amsfonts}
\usepackage{algorithmic}
\usepackage{textcomp}
\usepackage{derivative}
\usepackage{color}
\usepackage[dvips]{epsfig}
\begin{document}

\title{Complex dynamics of a heterogeneous network of Hindmarsh-Rose neurons}

\titlerunning{Complex dynamics of a heterogeneous network of Hindmarsh-Rose neurons}        

\date{Received: date / Accepted: date}

\author{Zeric Tabekoueng Njitacke \and Sishu Shankar Muni \and Soumyajit Seth \and Jan Awrejcewicz \and Jacques Kengne}

\authorrunning{Njitacke et \textit{al.}} 

\institute{Z.T. Njitacke \at
              Department of Electrical and Electronic Engineering, College of Technology (COT), University of Buea, P.O. Box 63,Buea, Cameroon, Department of Automation, Biomechanics, and Mechatronics, Faculty of Mechanical engineering, Lodz University of Technology, Poland. \\
            \email{zerictabekoueng@yahoo.fr}\\
             S.S. Muni \at
              School of Fundamental Sciences, Massey University, Palmerston North, New Zealand \\
            \email{s.muni@massey.ac.nz} \\            
                        S. Seth \at
              Department of Automation, Biomechanics, and Mechatronics, Faculty of Mechanical engineering, Lodz University of Technology, Poland. \\
            \email{soumyajit.seth@p.lodz.pl} \\
            J. Awrejcewicz \at
              Department of Automation, Biomechanics, and Mechatronics, Faculty of Mechanical engineering, Lodz University of Technology, Poland. \\
            \email{jan.awrejcewicz@p.lodz.pl}\\
                       J. Kengne \at
              Research Unit of Automation and Applied Computer (URAIA), Electrical Engineering Department of IUT-FV,
            University of Dschang, P.O. Box 134, Bandjoun, Cameroon. \\
            \email{kengnemozart@yahoo.fr}
}

\maketitle

\begin{abstract}
In this contribution, we have considered the collective behavior of two as well as the network of heterogeneous coupled Hindmarsh–Rose (HR) neurons. The heterogeneous models were made of a memristive $2$-D (HR) and the traditional $3$-D HR neurons. Investigating a model of two coupled neurons through an electrical synapse reveals dissipative properties. When control parameters are varied, the coupled neuron model exhibits rich dynamics, such as the periodic, quasi-periodic, and chaotic dynamics involving either bursting or spiking oscillations. For weak electrical coupling strength, non-synchronized motion is observed. But in the case of higher coupling strength, synchronized cluster states are observed. Besides, ring-star networks of up to $100$, under three different heterogeneous topologies are being investigated, and various spatiotemporal patterns are explored. It is found that the spatiotemporal patterns depend on the topology of the heterogeneous network considered. A new clustered chimera state is revealed qualitatively via the recurrence plot. The cluster states are indicated in the ring and star configurations of the heterogeneous network. Single and double-well chimera states have been revealed in the ring and ring-star structures. Finally, an equivalent electronic circuit for the two coupled heterogeneous is designed and investigated in the PSIM simulation environment. A perfect match is observed between the results obtained from the designed analog circuit and the mathematical model of the two coupled neurons, which supports the fact that our obtained results are not related to an artifact.

\keywords{Heterogeneous Hindmarsh–Rose Neurons \and Electrical Synapse  \and Ring-Star network \and Chimera State \and Cluster State \and Analog Electronic Circuit.}
\end{abstract}

\section{Introduction}
Communications between many neurons and between neurons and cells are realized in a specialized contact region, called a synapse. The number of synapses in the human nervous system is estimated to be around $10^{14}$ ($100$ trillion), but given newly discovered synaptic diversity, it could be substantially greater \cite{ref1}. Several research studies have been carried out to study the synapse based on artificial mathematical models. Among others, we can quote the chemical synapse \cite{ref2}, electrical synapse\cite{ref3, ref4}, hybrid synapse \cite{ref5, ref6}, Josephson junction synapse \cite{ref7}, and memristive synapse \cite{ref8, ref9}. These varieties of artificial synapses are used to couple artificial models of neurons. Among the artificial models of neurons proposed in the literature, the Hopfield neural network model \cite{ref10, ref11, ref12, ref13}, the Hodgkin–Huxley neuron \cite{ref14}, the Chay model \cite{ref15},  the Izhikevich neuron \cite{ref16}, the FitzHugh–Nagumo (FN) model \cite{ref17}, the Morris–Lecar neuron \cite{ref18}, the $2$-D Hindmarsh–Rose (HR), the $3$-D-HR model \cite{ref19, ref20},  the Rulkov model \cite{ref21} and the Wilson neuron \cite{ref22} are widely used.  For example, the dynamical behavior of a single non-autonomous Hopfield neuron with a memristive self-synaptic connection was studied in \cite{ref23}. They demonstrated that the proposed model might produce homogenous extreme multistability. They proposed a non-invasive control mechanism that allows them to choose any attractor among the coexisting ones, which is very interesting to observe. Telksnys et al. \cite{ref24} investigated the phenomena of broken symmetry in the single solutions of the Hodgkin–Huxley neuron model. They also showed the necessary and sufficient conditions for bright and dark solitary solutions in that model. Xu et al. \cite{ref22} predicted the dynamics of an improved Wilson neuron. They have considered a memristor model to imitate the effect of electromagnetic induction on it. As a result, rich electrical activities, including the asymmetric coexisting electrical activities and the anti-monotonicity phenomenon, were recorded in their model. The global dynamics of a Chay neuron was explored in \cite{ref25}. The authors demonstrated the occurrence of some neuron characteristics, such as bursting and spiking, using several basic nonlinear analytic methods. More importantly, their proposed neuron model was implemented on an FPGA to validate their theoretical investigations. The authors explored a Fitzhugh–Nagumo neuron model with memristive autapse in \cite{ref26}. According to their findings, the proposed model can display extreme multistability. Finally, the circuit implementation of the examined model supported their findings. The multistable dynamics of an autonomous Morris–Lecar neuron have been addressed in \cite{ref27}. During their study, diverse neuronal behaviors such as chaotic bursting, chaotic tonic-spiking, and periodic bursting behaviors were found. Also, the results were further validated based on a microcontroller development board. A model of a $4$-D memristive Hindmarsh–Rose has been investigated in \cite{ref28}.

Apart from the dynamical behavior, several researchers have studied the collective behavior of the network of neurons. In \cite{ref29}, the authors considered a discretized version of the Izhikevich neuron model and found that the electromagnetic flux can act as an order parameter in the sense that it can tune different firing patterns under the variation of electromagnetic flux. Hussain et al. \cite{ref30} investigated the dynamics of a network of multi-weighted Fitzhugh-Nagumo neurons, taking into account the effect of electrical, chemical, and ephaptic couplings. They analyzed the impact of those coupling on the chimera states and the complete synchronization exhibited by the networks. When the temperature coefficient was varied, chimera states were able to occur in a network of thermosensitive Fitzhugh-Nagumo neurons \cite{ref31}. In \cite{ref32}, the authors investigated a Hindmarsh-Rose neuronal network. They found the interconnected neurons were able to exhibit chimera states. They also investigated the effect of $\alpha-stable$ noise. They found that when the control parameter of the noise is increased, the chimera states are progressively suppressed. In \cite{ref33}, considering the impacts of electrical synapses and autapses, the authors investigated the behavior of a network of Hindmarsh-Rose neurons. They demonstrated the presence of chimera states in the studied network using numerical simulations. More interestingly, Simo et al. \cite{ref34} revealed traveling chimera patterns in a $2$-D HR neural network. Simo et al. \cite{ref35} studied a set of $3$-D HR neurons in the presence of an electric field. As a result, traveling chimera states and multicluster oscillating breathers were discovered in the absence of the field. In the presence of the field, chimera states, multi-chimera states, alternating chimera states, and multicluster traveling chimeras were discovered. In a ring network of bistable systems, like FitzHugh Nagumo neurons \cite{Sh17a}, Chua circuits \cite{Mu20a}, double-well chimera states were revealed. It was conjectured that the mechanism behind the formation of double-well chimera states was due to double-scroll chaotic attractors or even attractors that span both positive and negative values of the state variable. Interestingly, double-well chimera states were found in the current heterogeneous network under study.

In all of the preceding works, great emphasis has been given to investigating single neuron and the networks of neurons. Concerning networks of neurons, most studies generally considered only networks with the same types of neurons. Such networks are well known to be homogeneous. In a nutshell, the homogeneous coupled neurons or networks showed the complete synchronization, phase synchronization, chimera states, multi-chimera states, alternating chimera states, and multicluster traveling chimeras. In contrast, the dominant behaviors observed in our heterogeneous coupled neurons network were robust chaos, clusters of synchronization, and clustered chimera state—single and double-well chimera states. It motivates us to consider the heterogeneous coupling between HR neuron models, which will be addressed with the following objectives:
\begin{itemize}
\item Investigate the global behavior of a $2$-D mHR neuron coupled with a $3$-D HR via a memristive synapse.
\item Investigate the network of up to $100$ neurons using the various topological configuration of the $2$D mHR and the 3D HR neurons.
\item Propose an electronic circuit of the coupled two neurons to validate our numerical results.
\end{itemize}
This article has been structured as follows. In Section~\ref{md}, we have described the considered model, which shows the heterogeneous interactions, and have studied their properties. Section~\ref{db} shows the dynamical behaviors of the supposed system, as a transition from periodicities to co-existence of different attractors, etc. In section~\ref{na}, we have analyzed the networks of oscillators. Section~\ref{expt_results} depicts the experimental validations of the numerical predictions of the two coupled neuron models. Section~\ref{conclusion} is the conclusion of our work.

\section{Model description and its properties}
\label{md}

\subsection{Model description}
\begin{equation}\label{eq1}
\left\{ \begin{array}{l}
 \dot x_1  = y_1  - a_1 x_1^3  + b_1 x_1^2  + i_1  + \alpha \cos \left( {z_1 } \right)x_1  \\ 
 \dot y_1  = c_1  - d_1 x_1^2  - y_1  \\ 
 \dot z_1  = \sin \left( {z_1 } \right) + ex_1  \\ 
 \end{array} \right.
\end{equation}
\begin{equation}\label{eq2}
\left\{ \begin{array}{l}
 \dot x_2  = y_2  - a_2 x_2^3  + b_2 x_2^2  - z_2  + i_2  \\ 
 \dot y_2  = c_2  - d_2 x_2^2  - y_2  \\ 
 \dot z_2  = r\left( {s\left( {x_2  - \bar x_2 } \right) - z_2 } \right) \\ 
 \end{array} \right.
\end{equation}
\begin{equation}\label{eq3}
\left\{ \begin{array}{l}
 \dot x_1  = y_1  - a_1 x_1^3  + b_1 x_1^2  + i_1  + \alpha \cos \left( {z_1 } \right)x_1  + \sigma \left( {x_2  - x_1 } \right) \\ 
 \dot y_1  = c_1  - d_1 x_1^2  - y_1  \\ 
 \dot z_1  = \sin \left( {z_1 } \right) + ex_1  \\ 
 \dot x_2  = y_2  - a_2 x_2^3  + b_2 x_2^2  - z_2  + i_2  - \sigma \left( {x_2  - x_1 } \right) \\ 
 \dot y_2  = c_2  - d_2 x_2^2  - y_2  \\ 
 \dot z_2  = r\left( {s\left( {x_2  - \bar x_2 } \right) - z_2 } \right) \\ 
 \end{array} \right.
\end{equation}
A brain-like complex structure is made of the interconnection of many neurons. Those neurons have a variety of functionality and thus perform many tasks. Most studies in the literature are generally focused on the networks of neurons that are homogeneous under the effect of different types of fields. On the other hand, these works are the ideal examples because they have concentrated on homogeneous networks of neurons to explain complex brain behavior. To consider a more realistic system and for more robust applications, we have proposed a model in the Eq.~(\ref{eq3}) consisting of the heterogeneous coupled neurons. They are constructed with a memristive $2$-D HR neuron (as shown in Eq.~(\ref{eq1})) associated with a traditional $3$-D HR neuron shown in Eq.~(\ref{eq2}). The dynamics of the coupled model are investigated using standard nonlinear analysis methods in the parameter space, such as one parameter bifurcation diagrams, maximal Lyapunov exponent spectrums, and the two-parameter bifurcation diagrams. Also, a heterogeneous ring-star network of up to $100$ neurons is studied, and their collective dynamics are widely investigated.

In Eq. (\ref{eq3}), the two neurons are coupled through an electrical synapse having a coupling strength, $\sigma$ . An electrical synapse, also known as a gap junction, is a mechanical contact between two neurons that permits electricity to pass through. The channels in electrical synapses allow charges (or ions) to flow from one cell to another. In this way, the neurons can exchange informations and communicate with millions of others via the complex chain reactions within interconnected neurons. In Eq.~(\ref{eq3}), $x_{i}$s  are the membrane potentials of each HR neuron, also called fast variables, $y_{i}$s are the retrieval variables related to a fast current of either $Na^{+}$ or $K^{+}$, which is also called recovery variables. The state variable $z_{1}$ of the $2$-D memristive HR neuron stands for the inner variable of the memristive autapse. The term $sin(z_{1}) + ex_{1}$ represents the superposition of the magnetic flux leakage, and the membrane potential enabling variation on magnet flux, respectively. The term $\alpha cos(z_{1}) x_{1}$ stands for the memductance of the memristive synapse. It shows the modulation of time-varying field on the gap junction of the membrane. The state variable, $z_{2}$, is the slow adaptation current of the traditional $3$-D HR neuron, that represent external forcing currents while $\alpha$ represents the memristive autapse strength. The investigation of the collective behavior of the proposed coupled neurons and the network will be based on the following values of the parameters: $a_1  = a_2  = 1$, $b_1  = b_2  = 3$, $c_1  = c_2  = 1$, $d_1 = d_2  = 5$, $r = 0.008$, $\bar x_2  = - 1.6$, $s = 4$, $e = 0.5$, $f = 0.5$, $i_1  = m\sin \left({2\pi ft} \right)$. The parameters $\alpha$ and $i_{2}$ are tuneable parameters for different values of $\sigma$, $m$.

\subsection{Dissipation properties}
\begin{gather}\label{eq4}
\small
\nabla.V =
 \begin{bmatrix}
 \pdv{}{x_1} \\  \pdv{}{y_1} \\  \pdv{}{z_1} \\  \pdv{}{x_2} \\  \pdv{}{y_2} \\  \pdv{}{z_2}
 \end{bmatrix}
 .
 \begin{bmatrix}
  y_1 + b_1x_1^2 - a_1x_1^3  + i_1 + \alpha x_1 \cos(z_1) + \sigma (x_2 - x_1) \\
  c_1 - d_1x_1^2 - y_1 \\
  ex_1 + \sin(z_1) \\
  y_2 + b_2x_2^2 - a_2x_2^3  + i_2 - z_2 - \sigma (x_2 - x_1) \\
  c_2 - d_2x_2^2 - y_2 \\
  r\big(s(x_2 - \bar{x}_2) - z_2\big) \\
  \end{bmatrix}
 \end{gather}
 \begin{equation}\label{eq5}
 \begin{split}
\nabla.V{\rm{ }} = &  - \left( {3a_1 \left( {x_1^2  + x_2^2 } \right) + 2\sigma  + 2 + r} \right) + 2b_1 \left( {x_1  + x_2 } \right) \\& + \cos \left( {z_1 } \right)\left( {\alpha  + 1} \right)
\end{split}
\end{equation} 
The volume contraction rate of the heterogeneous coupled neurons considered in the first part of this work is given in Eq.~(\ref{eq4}). The contraction rate enables us to estimate an ideal nature of firing patterns generated by the proposed model. In such a way, three types of patterns can be identified among which, the dissipative one with $\nabla.V < 0$, the conservative one with  $\nabla.V= 0$, and the repelled one with $\nabla.V > 0$ .

From the Eq.~(\ref{eq5}), it is evident that the volume contraction rate of the coupled neurons depends on the average values of the membrane potentials of both the neurons as well as the inner variable of the memristive autapse of the $2$-D HR neuron. Therefore, the coupled model will be dissipative if and only if
\begin{equation}\label{eq6}
\begin{split}
& \nabla.V < 0 \Rightarrow \left( {3a_1 \left( {x_1^2  + x_2^2 } \right) + 2\sigma  + 2 + r} \right) > 2b_1 \left( {x_1  + x_2 } \right) \\& + \cos \left( {z_1 } \right)\left( {\alpha  + 1} \right)
\end{split}
\end{equation}
\begin{figure}[tbh]
\centering
\begin{subfigure}[b]{0.49\linewidth}
\includegraphics[width=\linewidth]{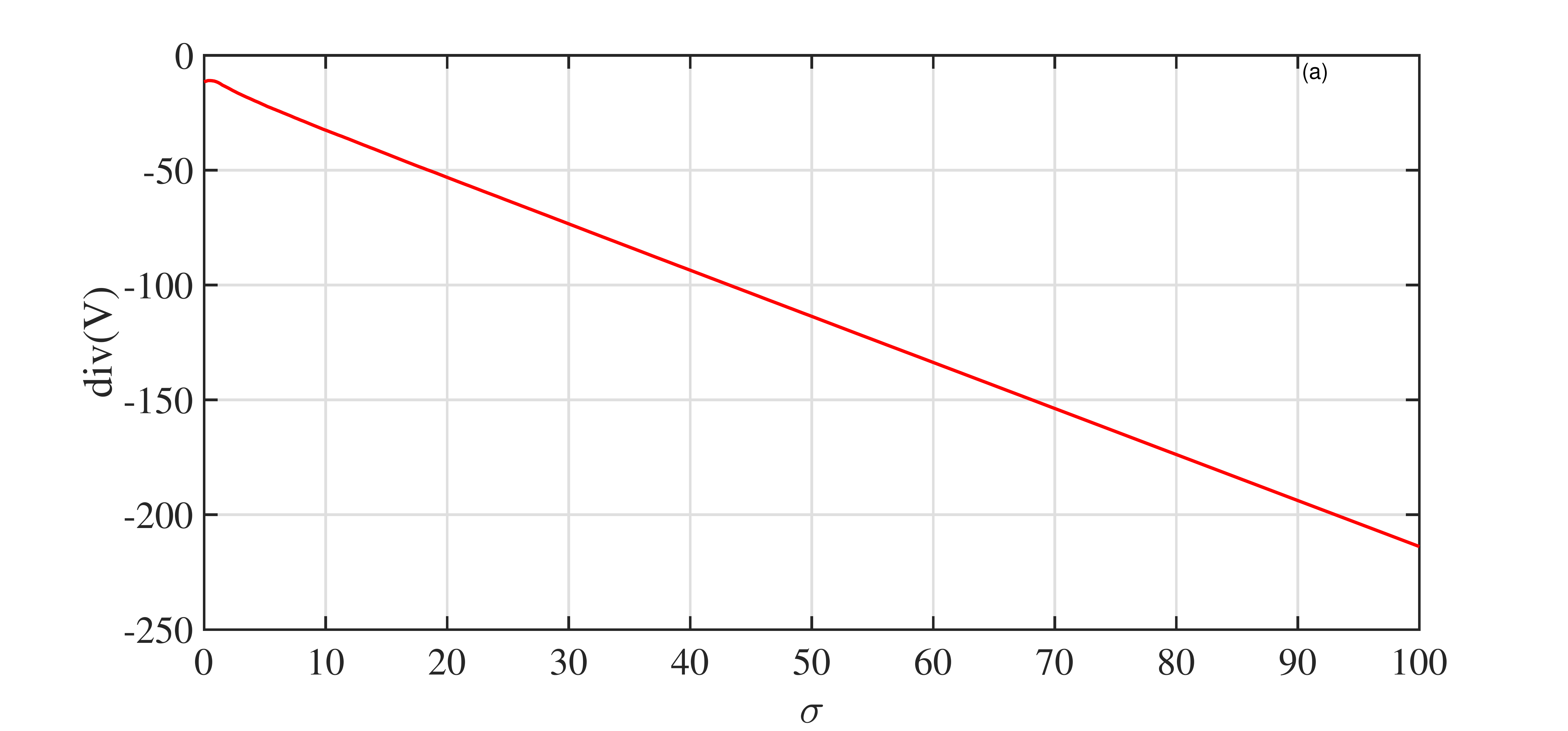}
\caption{}
\end{subfigure}
\begin{subfigure}[b]{0.49\linewidth}
\includegraphics[width=\linewidth]{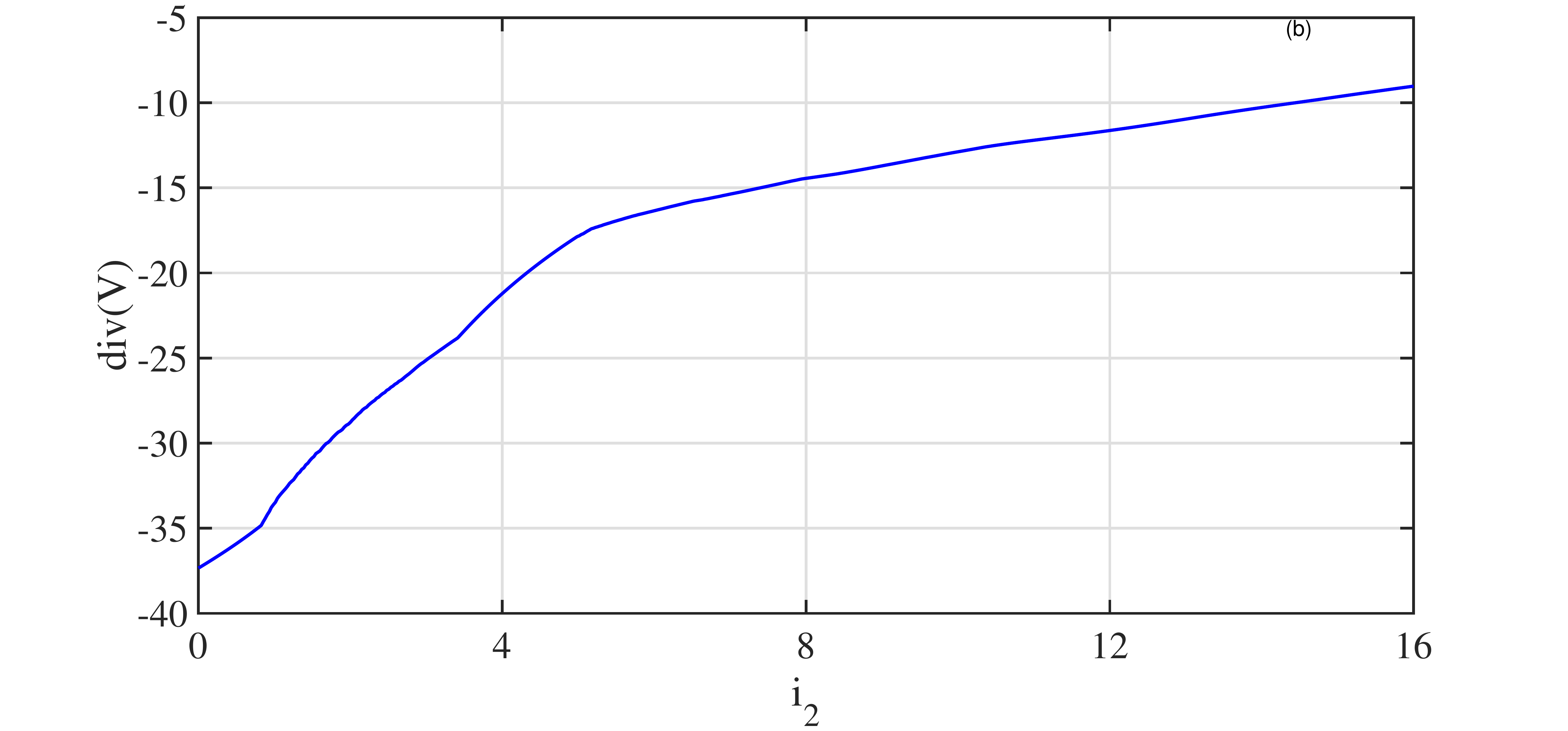}
\caption{}
\end{subfigure}
\caption{Evolution of the volume contraction rate of the coupled neurons when varying the control parameters (a) $\sigma$ and (b) $i_2$ from low to the high values in the positive directions. The remaining parameters are kept constant in the case of (a) at $ i_2$ = $3$, $m$ = $2$, $\alpha=0.5$ and in the case of (b) at $\sigma$ = $2$, $m$ = $2$, $\alpha$ = $0.5$.   The initial conditions for the both figures are chosen at ($-0.54$,$-5,0$,$0.1$,$0$,$0$,$6.688$). (Color online)}
\label{fig1} 
\end{figure}
The volume contraction rate of the Eq.~(\ref{eq5}) has been numerically computed by varying one of the control parameters, $\sigma$, keeping the remaining parameters to fixed values, $i_2=3$, $m = 2$, and $\alpha = 0.5$. It is shown in Fig.~\ref{fig1}(a). The same volume contraction rate has been computed again when the parameter $i_2$ is smoothly varied, and the rest parameters to be fixed at $\sigma=2$, $m = 2$, and $\alpha = 0.5$, as depicted in Fig.~\ref{fig1}(b). Therefore, it is obvious to conclude that for any set of parameter values used to compute Fig.~\ref{fig1}, the volume contraction rate of the coupled neurons is always negative. Consequently, the considered model is dissipative and can support the presence of attractors.

\section{Dynamical behavior of  the coupled neurons}
\label{db}
\begin{figure}[tbh]
\centering
\begin{subfigure}[b]{0.49\linewidth}
\includegraphics[width=\linewidth]{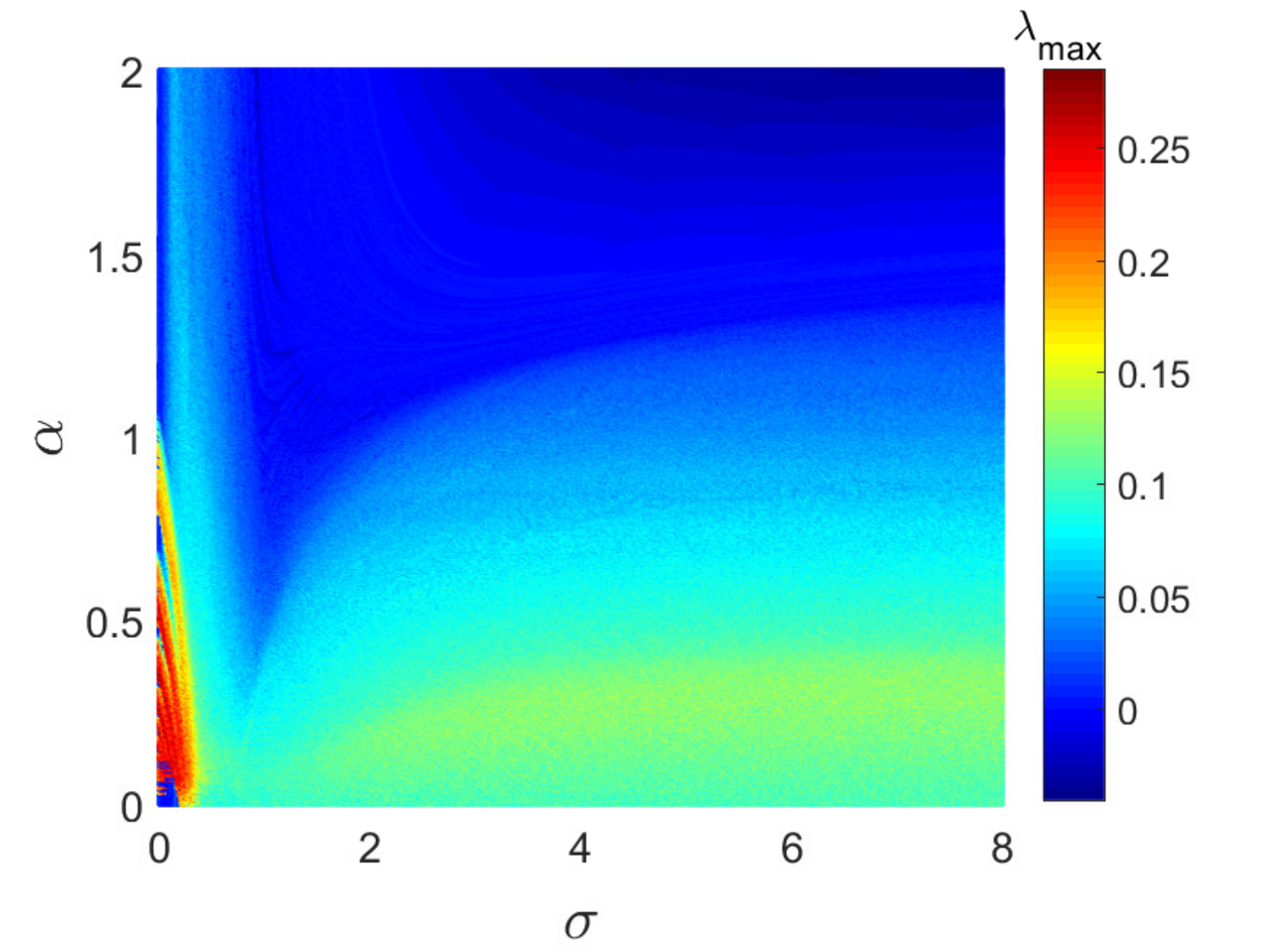}
\caption{}
\end{subfigure}
\begin{subfigure}[b]{0.49\linewidth}
\includegraphics[width=\linewidth]{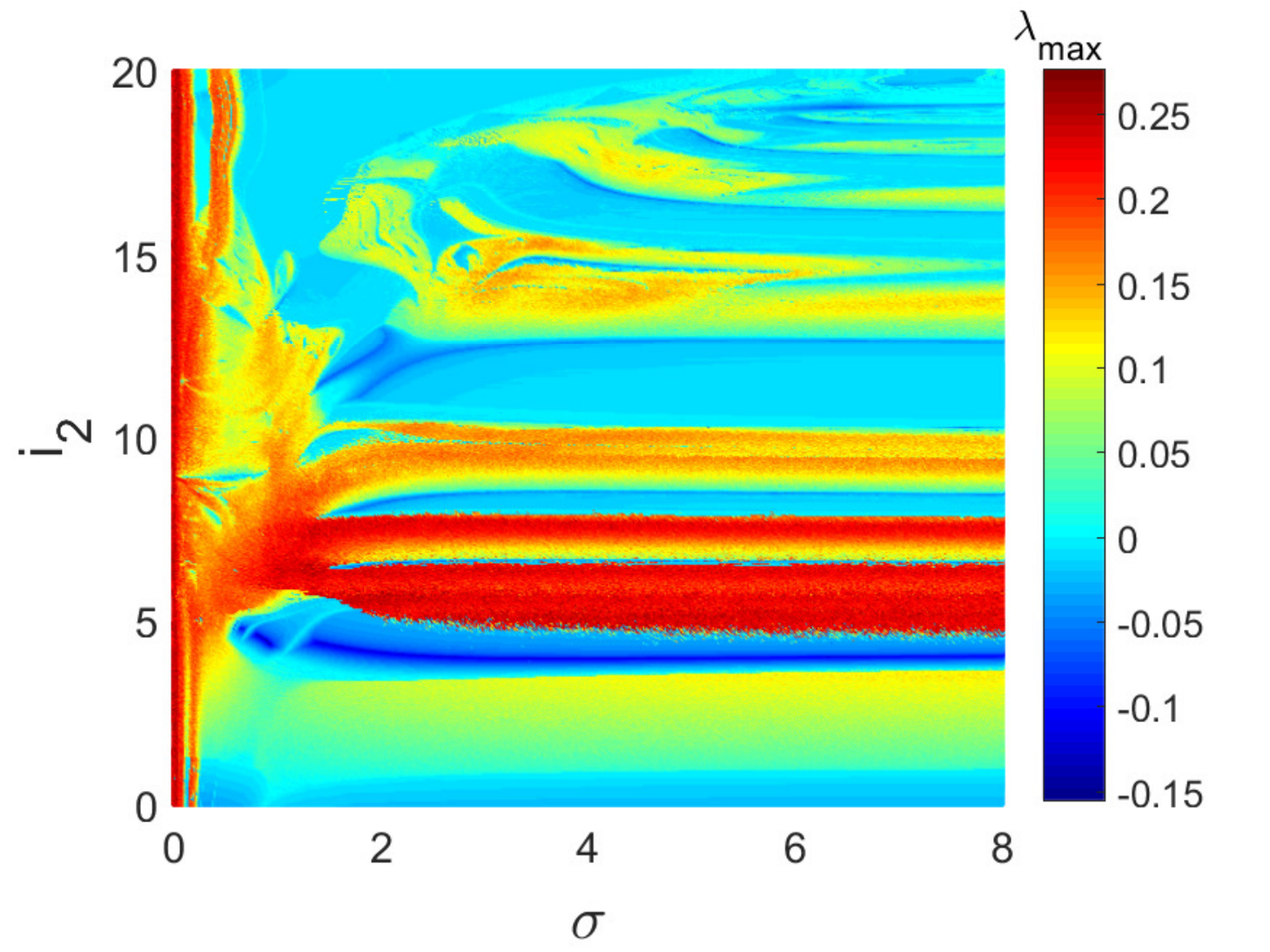}
\caption{}
\end{subfigure}
\begin{subfigure}[b]{0.49\linewidth}
\includegraphics[width=\linewidth]{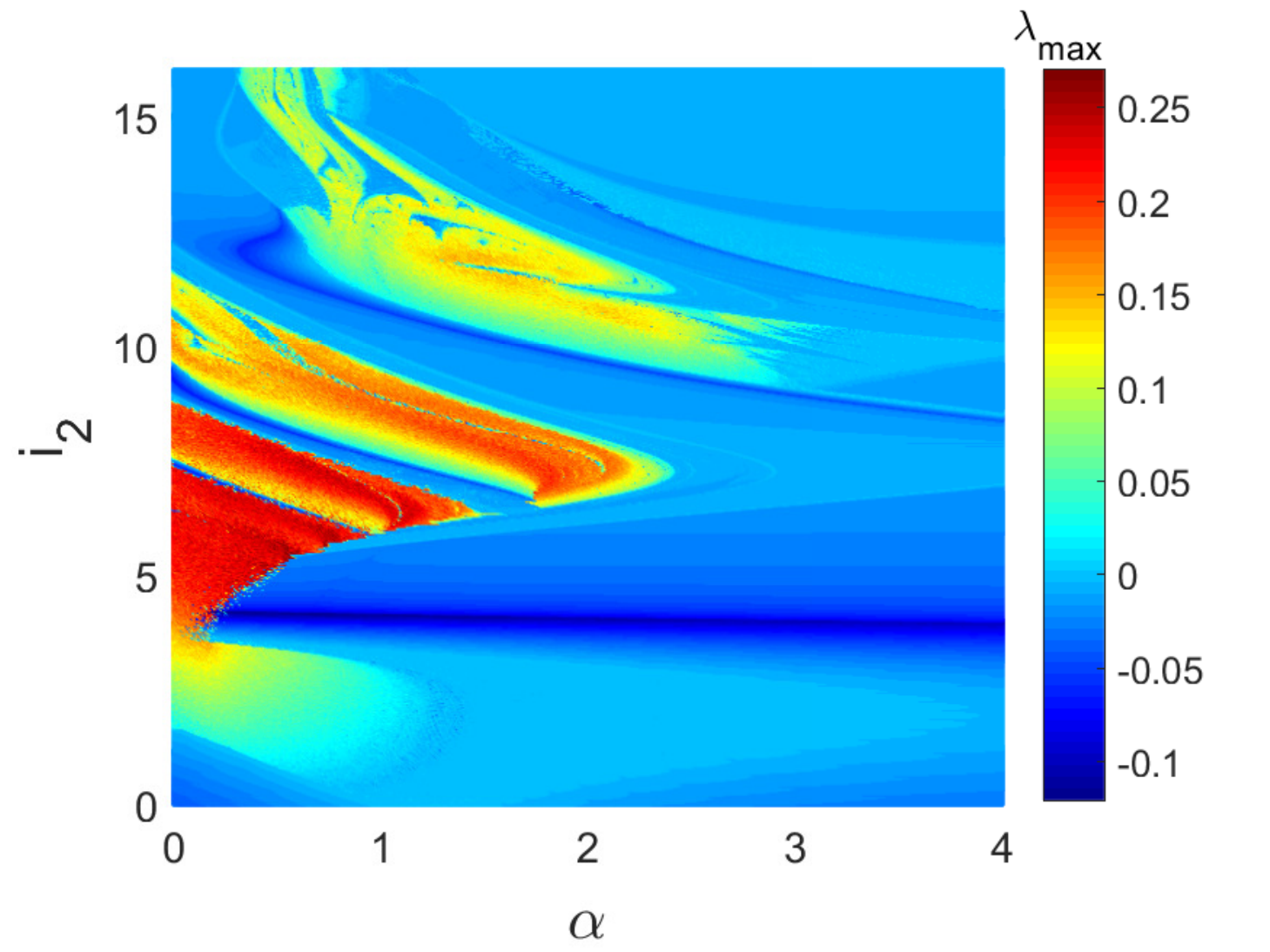}
\caption{}
\end{subfigure}
\begin{subfigure}[b]{0.49\linewidth}
\includegraphics[width=\linewidth]{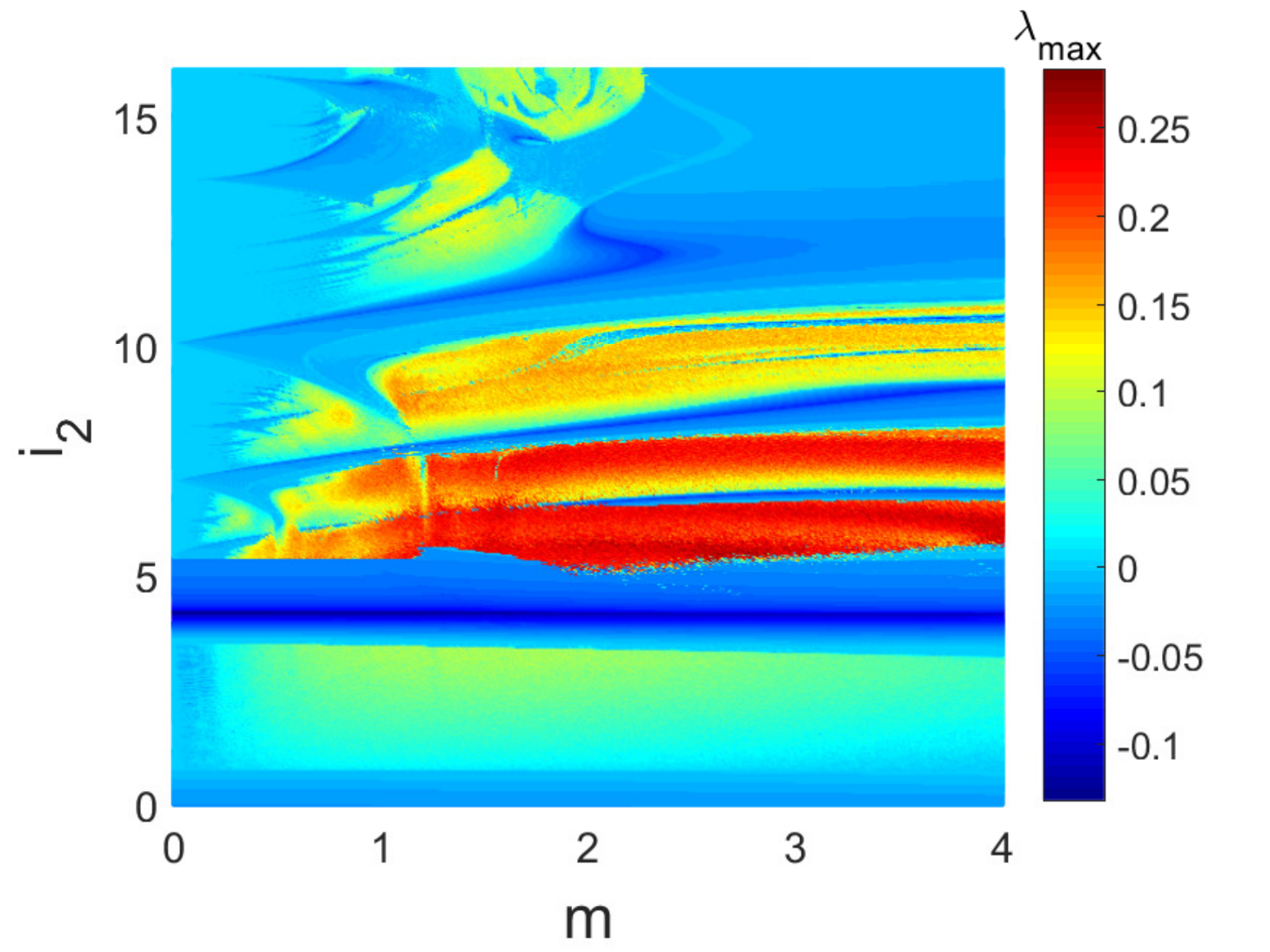}
\caption{}
\end{subfigure}
\medskip
\caption{Two parameters maximal Lyapunov exponent graphs. (a) The plane $(\sigma,\alpha)$ for $i_2 = 3$ and $m = 2$. (b) The plane $(\sigma,i_2)$ for $\alpha = 0.5$ and $m = 2$. (c) The plane $(\alpha,i_2)$ for $\sigma = 2$ and $m = 2$. (d) The plane $(m,i_2)$ for $\sigma = 2$ and $\alpha = 0.5$. For each of the figures, the initial conditions are ($-0.54$,$-5,0$,$0.1$,$0$,$0$,$6.688$). (Color online)}
\label{fig2} 
\end{figure}
This section is devoted to investigating the dynamical behavior of two coupled heterogeneous HR neuron models via an electrical synapse. The two-parameters maximal Lyapunov exponents, bifurcation diagrams, characterize the global behavior of the coupled neurons. In addition, numerical simulations have been performed using parameters and variables in extended precision mode with a fixed time step of $5 \times 10^{-3}$. When the two parameters of the coupled neurons are simultaneously varied, and the maximum Lyapunov exponent is recorded at each iteration, the color-coded figures (as shown in Fig.~\ref{fig2}) are obtained. The system's dynamics (such as periodic attractors, quasi periodicities, and chaos) are recorded from these two-parameters Lyapunov exponent diagrams. Periodic behaviors with the regular attractors are characterized by $\lambda_{\rm max} < 0$, the quasiperiodic  behaviors are supported by $\lambda_{\rm max} = 0$, while the chaotic behaviors with bounded random patterns are supported by $\lambda_{\rm max} > 0$. 

\begin{figure}[tbh]
\centering
\begin{subfigure}[b]{0.49\linewidth}\centering
\includegraphics[width=\linewidth]{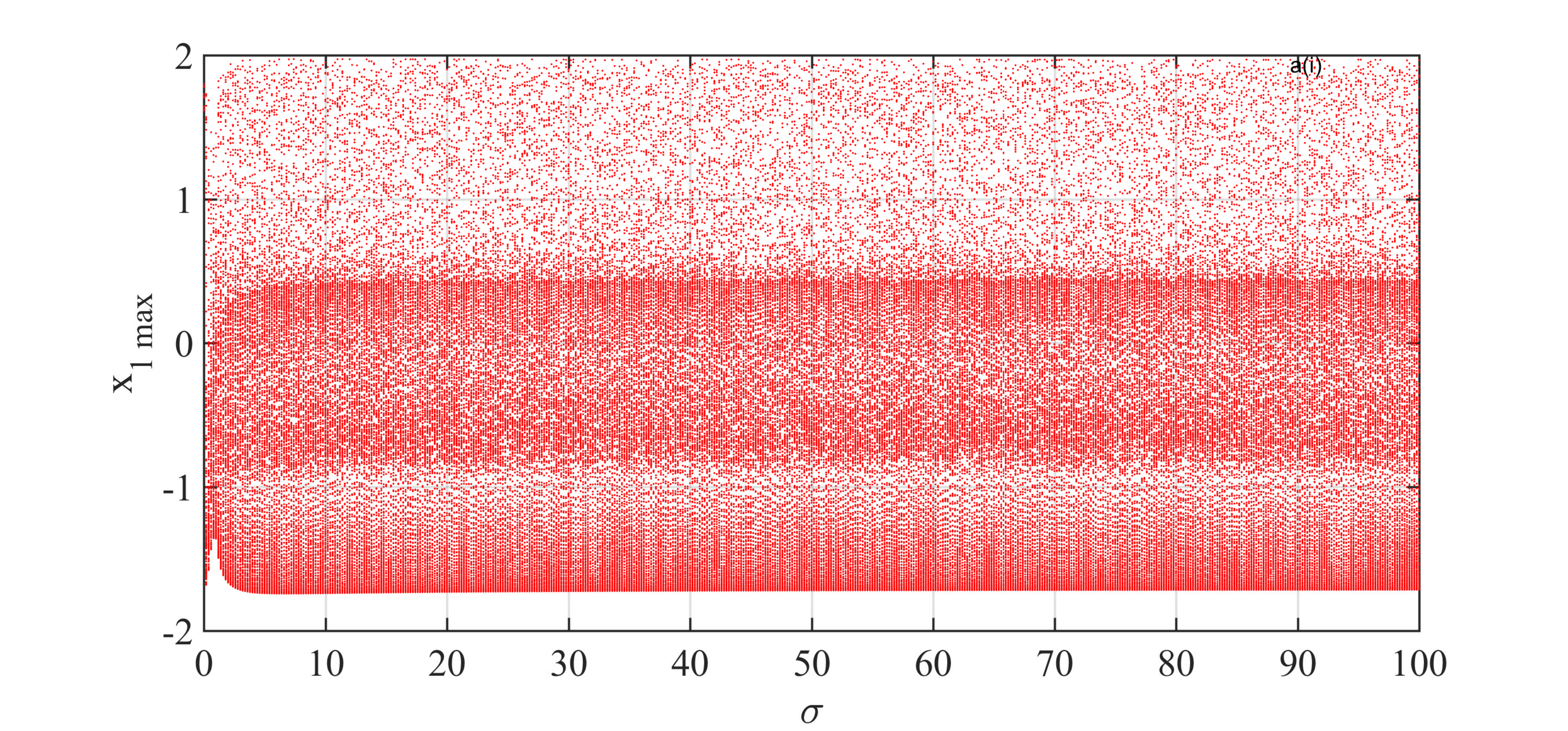}
\caption{}
\end{subfigure}
\begin{subfigure}[b]{0.49\linewidth}
\includegraphics[width=\linewidth]{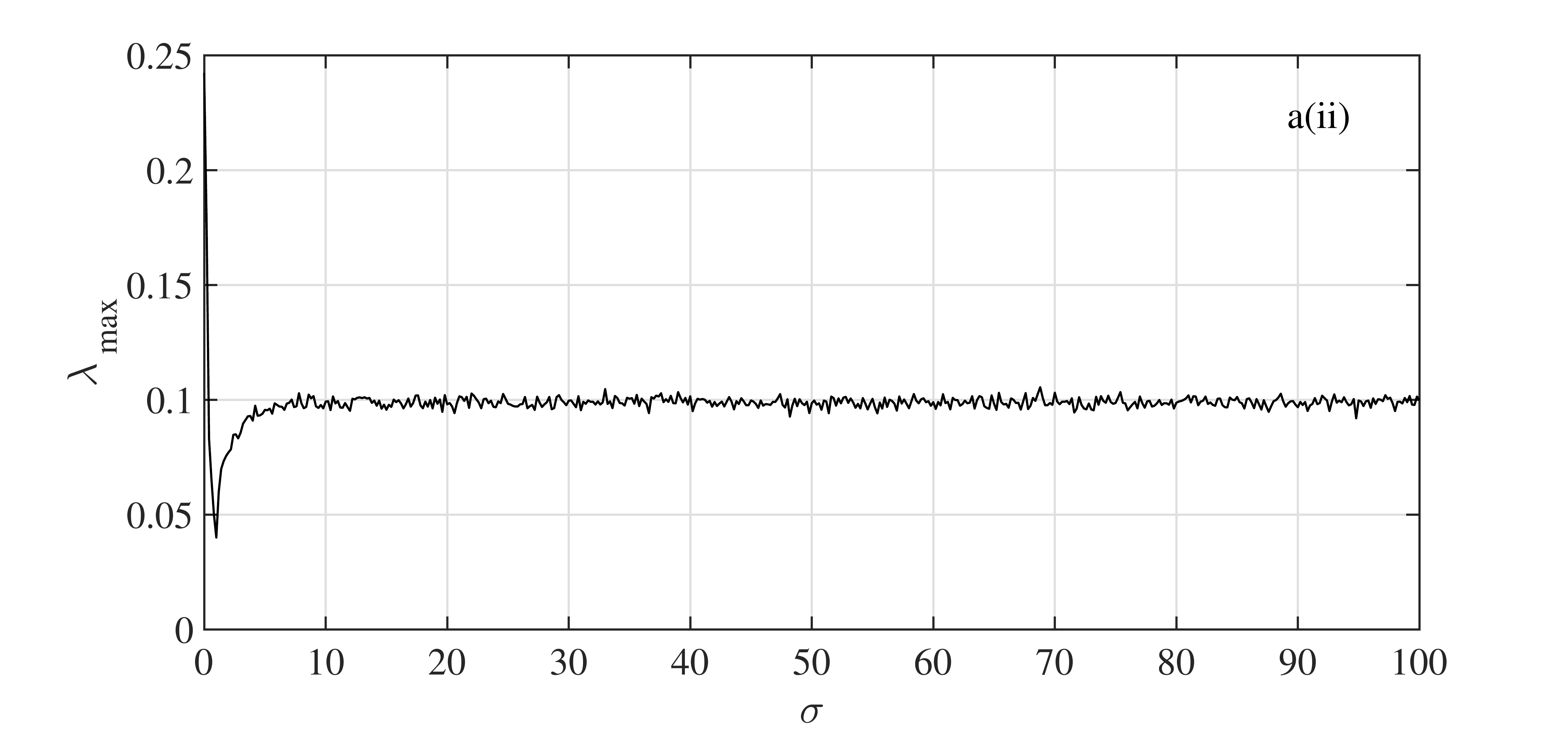}
\caption{}
\end{subfigure}
\begin{subfigure}[b]{0.49\linewidth}
\includegraphics[width=\linewidth]{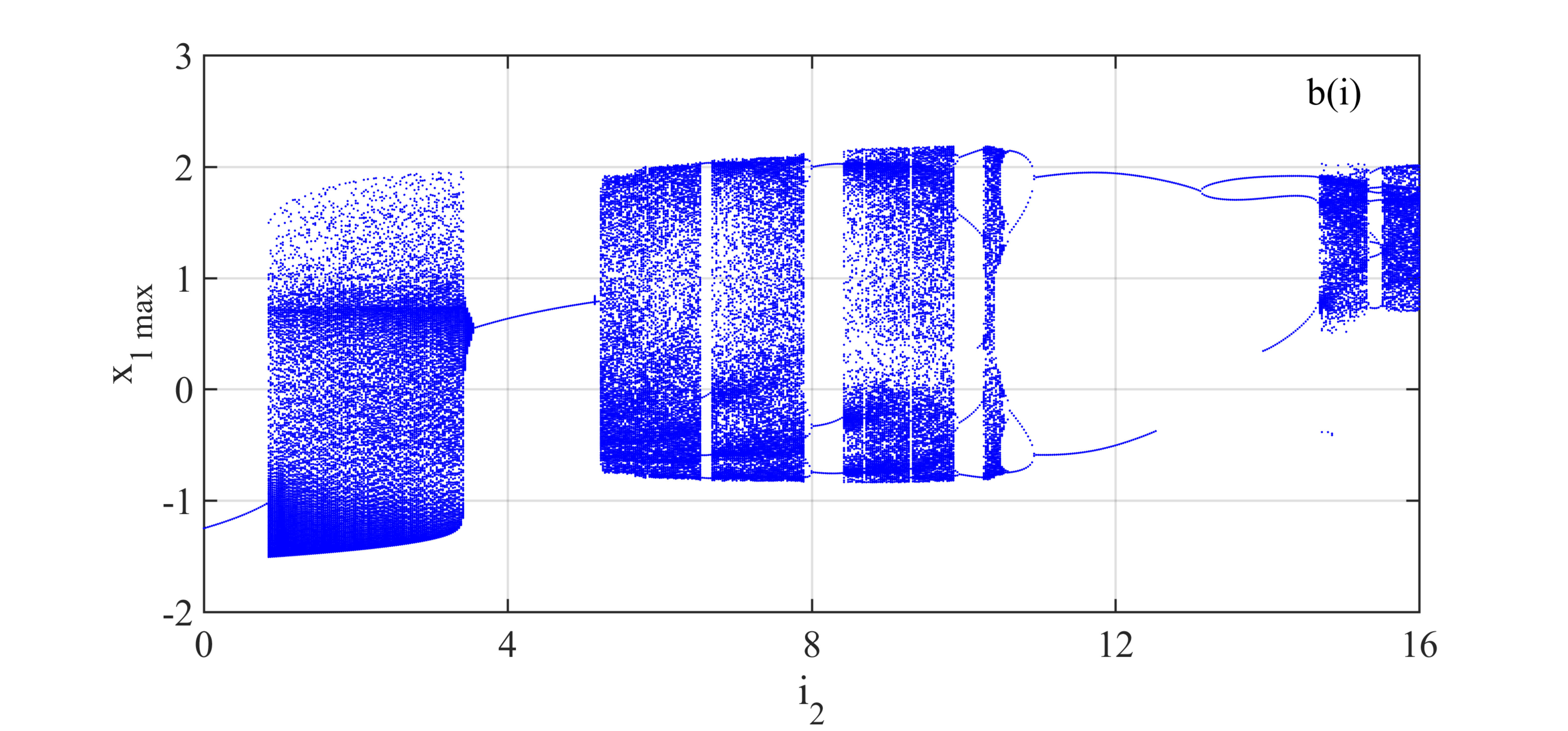}
\caption{}
\end{subfigure}
\begin{subfigure}[b]{0.49\linewidth}
\includegraphics[width=\linewidth]{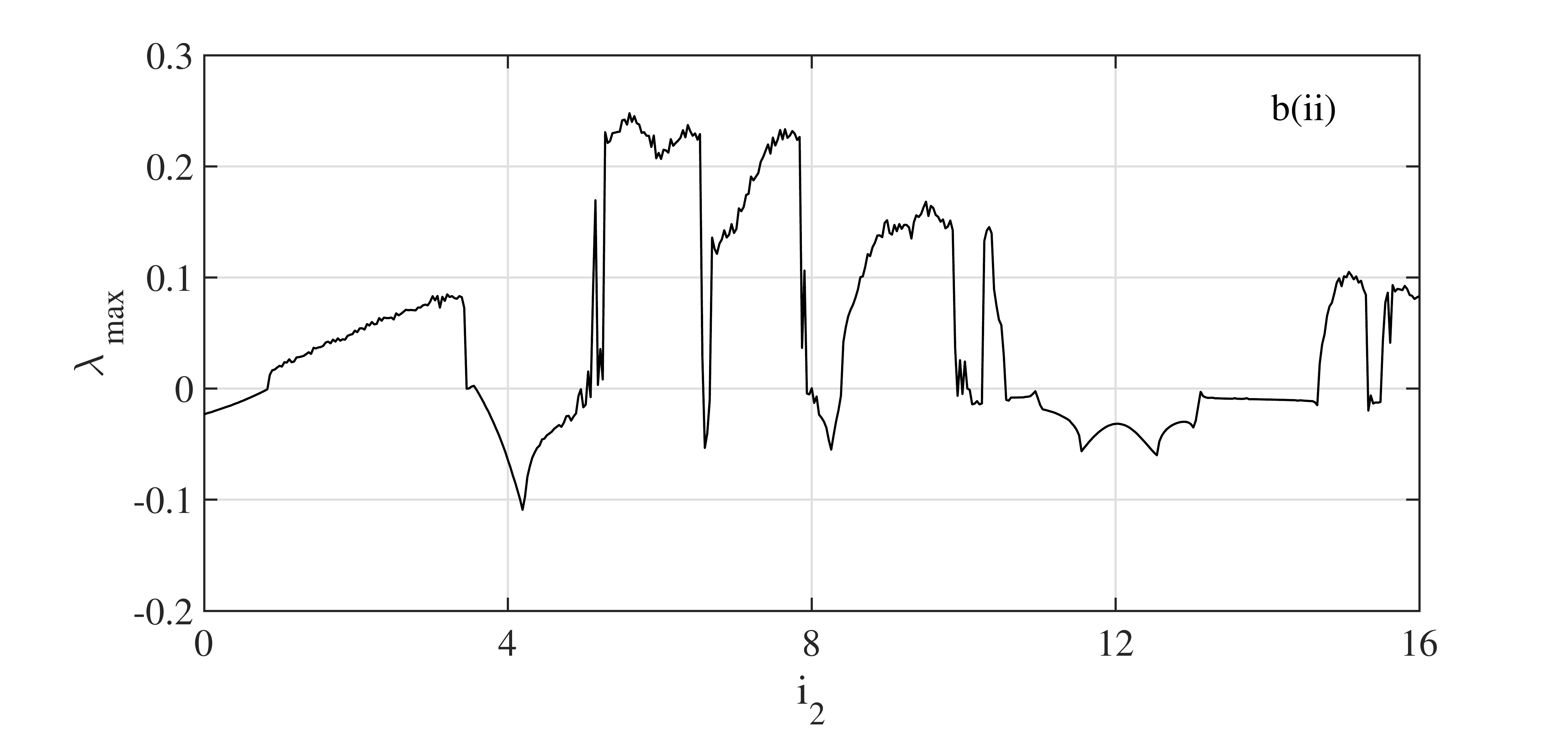}
\caption{}
\end{subfigure}
\begin{subfigure}[b]{0.49\linewidth}
\includegraphics[width=\linewidth]{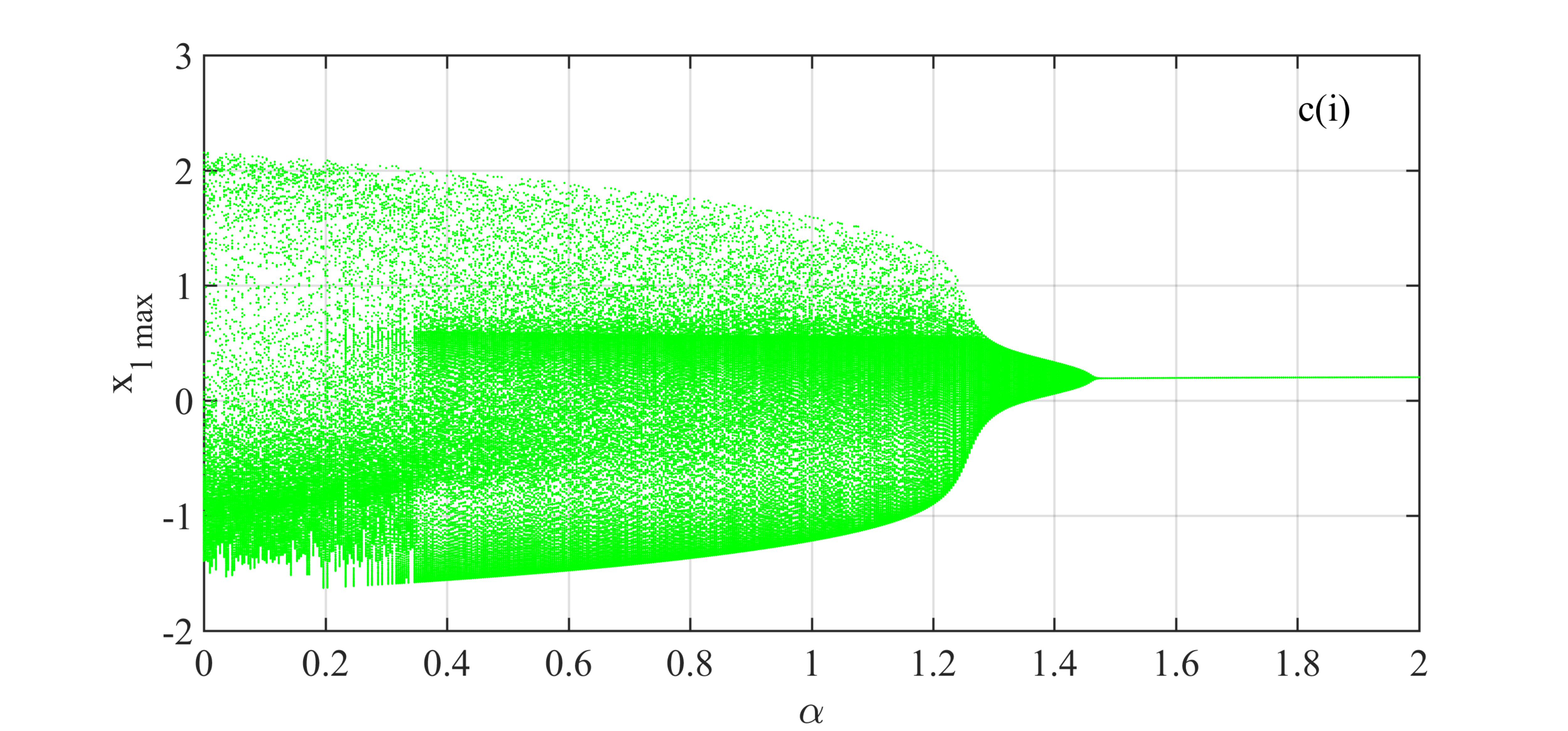}
\caption{}
\end{subfigure}
\begin{subfigure}[b]{0.49\linewidth}
\includegraphics[width=\linewidth]{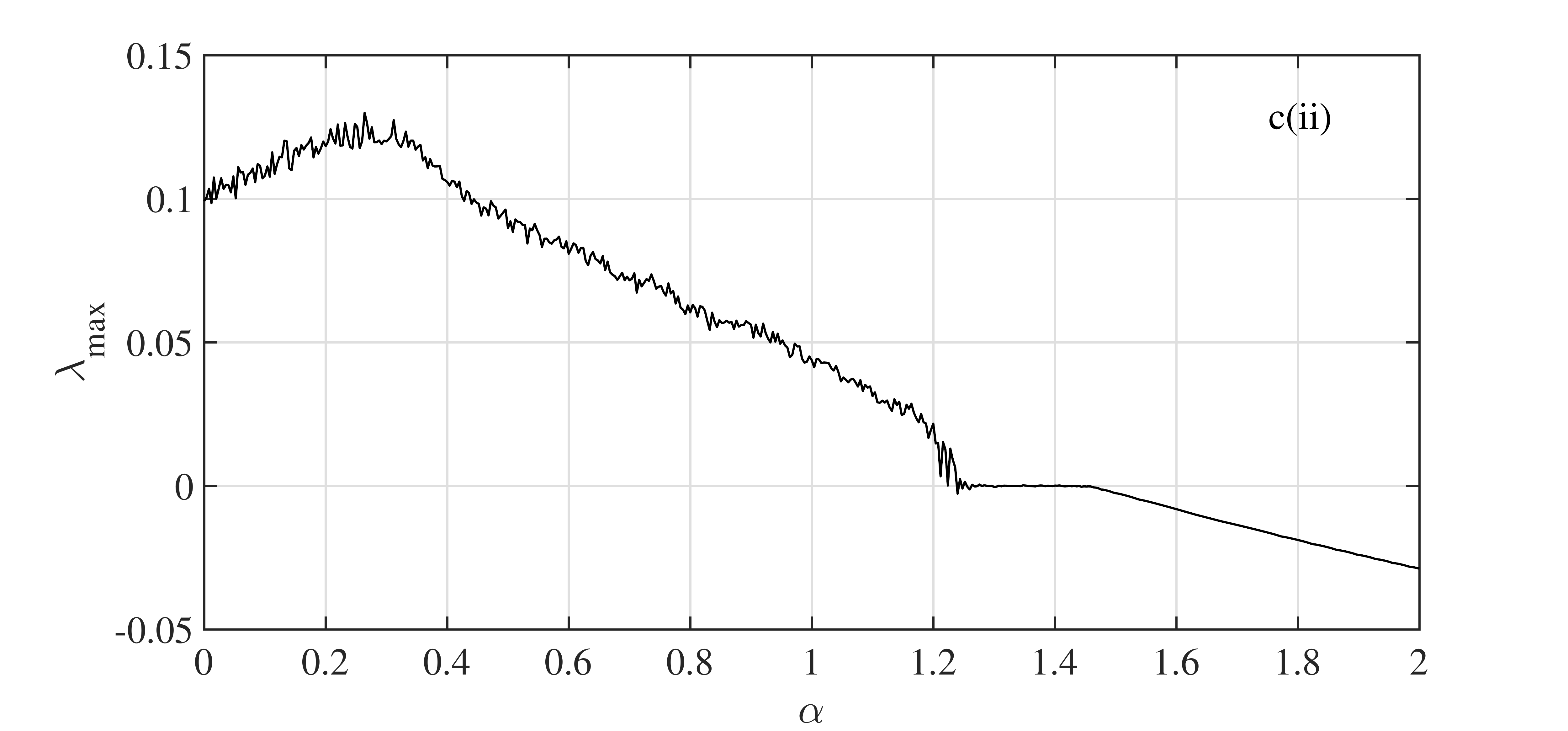}
\caption{}
\end{subfigure}
\medskip
\caption{The bifurcation diagrams and the corresponding maximal Lyapunov exponents with the variation of (a),(b) $\sigma$, (c),(d) $i_2$, (e),(f) $\alpha$. (a) and (b) are obtained for $ i_{2}=3, m=2 $, and $ \alpha=0.5 $. (c) and (d) are obtained for $ \sigma=2, m=2$, and  $\alpha=0.5 $. (e) and (f) are obtained for $ \sigma=4, m=2 $, and $ i_{2}=3$. For each of the figures, the initial conditions are ($-0.54$,$-5,0$,$0.1$,$0$,$0$,$6.688$). (Color online)}
\label{fig3} 
\end{figure}
It can be seen in the two-parameter diagrams that there are several windows of different attractors, like the periodic, quasiperiodic, and chaotic attractors. The switching between the attractors has been captured in the varying parameter space. In order to further support the dynamics of the coupled system, the bifurcation diagrams have been computed, which are shown in Fig.~\ref{fig3}(a), \ref{fig3}(c), and \ref{fig3}(e). The corresponding maximal Lyapunov exponents in the parameter space are shown in the Fig.~\ref{fig3}(b), \ref{fig3}(d), and \ref{fig3}(f), respectively. In the Fig.~\ref{fig3}(a), the bifurcation parameter is $\sigma$. We can see a transition from the quasiperiodic orbit to the chaotic orbit in the bifurcation diagram. The corresponding maximal Lyapunov exponent confirm this type of transition. The bifurcation diagram in Fig.~\ref{fig3}(c) is obtained when the external forcing current, $i_2$ of the second neuron, is varied. From the figure, we can see an interplay between different periodic orbits and the chaotic attractors in the bifurcation diagram. The periodic windows with different attractors exist between the chaotic attractors in the parameter space. The corresponding maximal Lyapunov exponent is shown in Fig.~\ref{fig3}(d).

Additionally, for some fixed values of the control parameters, the effects of the memristive autapse strength of the first neuron on the collective behavior of the coupled neurons are also evaluated. In the bifurcation diagram shown in Fig.~\ref{fig3}(e), when the memristive autapse $\alpha$ is null, the coupled model exhibits chaotic dynamics. When $\alpha$ is increased in the positive direction, the chaotic dynamic is metamorphosed to give birth to a quasiperiodic attractor. While expanding the control parameter $\alpha$ further, the quasiperiodic attractor disappears, and a period-1 limit cycle exists. The corresponding maximal Lyapunov exponent, as shown in Fig.~\ref{fig3}(f), confirms the transition between the attractors with the variation of $\alpha$.

The coupled neuron model, for example, can exhibit robust chaos when the electrical coupling strength $\sigma $ is varied within a specific range, as shown in  Fig.~\ref{fig3}(a). The persistence of chaotic dynamics in the coupled neurons model when the electrical coupling between the neurons is smoothly varied over a wide range characterizes this phenomenon \cite{banerjee1998robust, ref36, seth2019observation}.

\begin{figure}[tbh]
\centering
\begin{subfigure}[b]{0.49\linewidth}
\includegraphics[width=\linewidth]{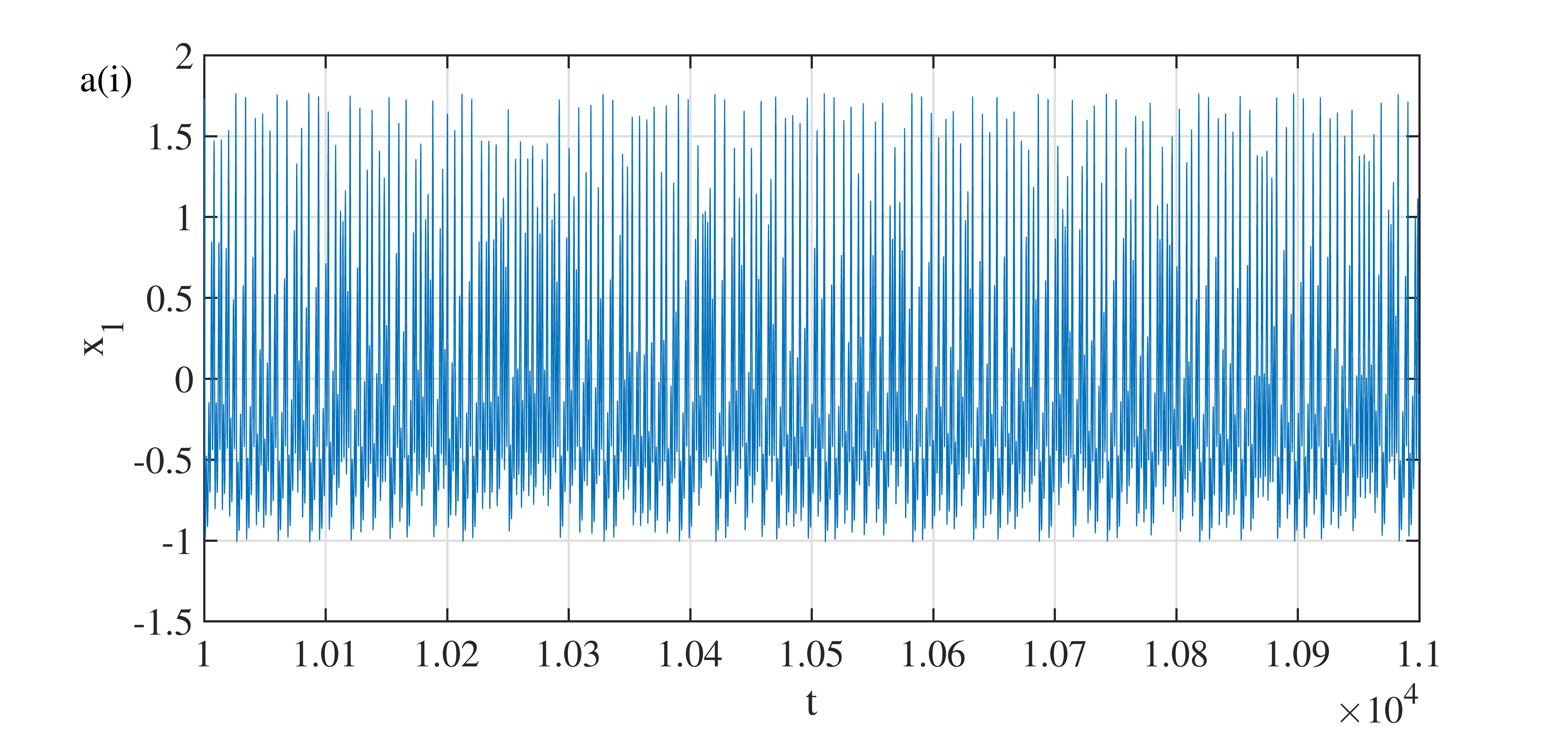}
\caption{}
\end{subfigure}
\begin{subfigure}[b]{0.49\linewidth}
\includegraphics[width=\linewidth]{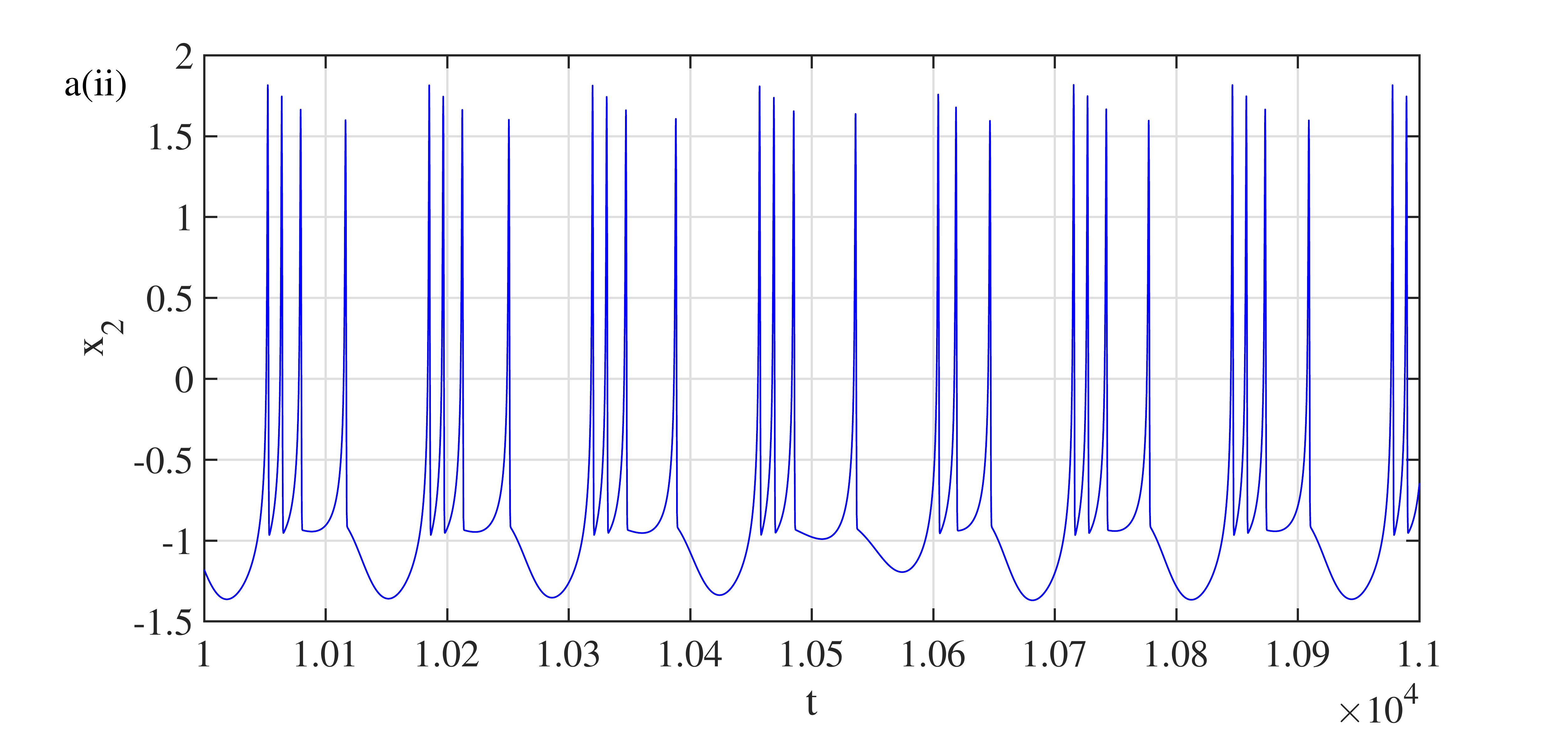}
\caption{}
\end{subfigure}
\begin{subfigure}[b]{0.49\linewidth}
\includegraphics[width=\linewidth]{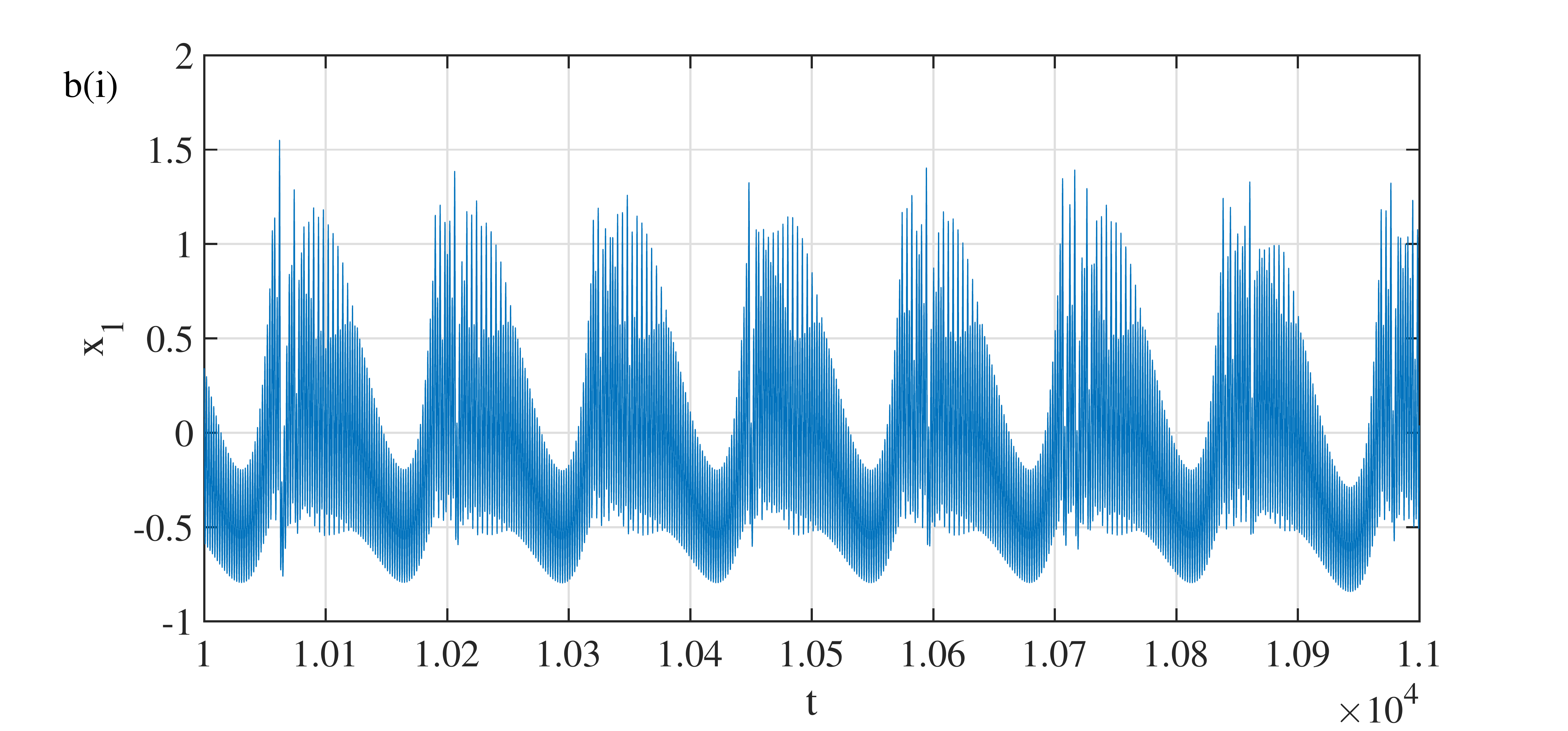}
\caption{}
\end{subfigure}
\begin{subfigure}[b]{0.49\linewidth}
\includegraphics[width=\linewidth]{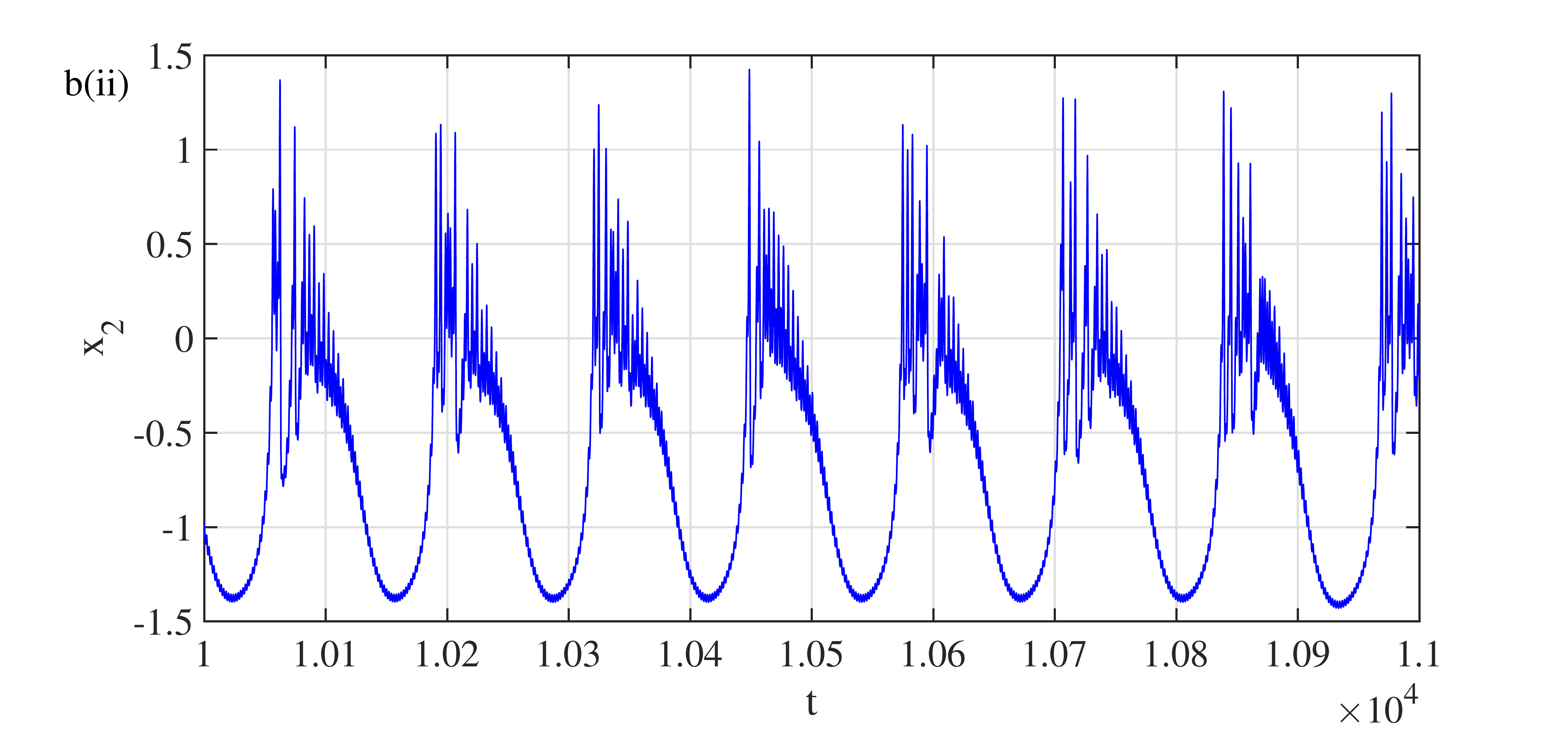}
\caption{}
\end{subfigure}
\begin{subfigure}[b]{0.49\linewidth}
\includegraphics[width=\linewidth]{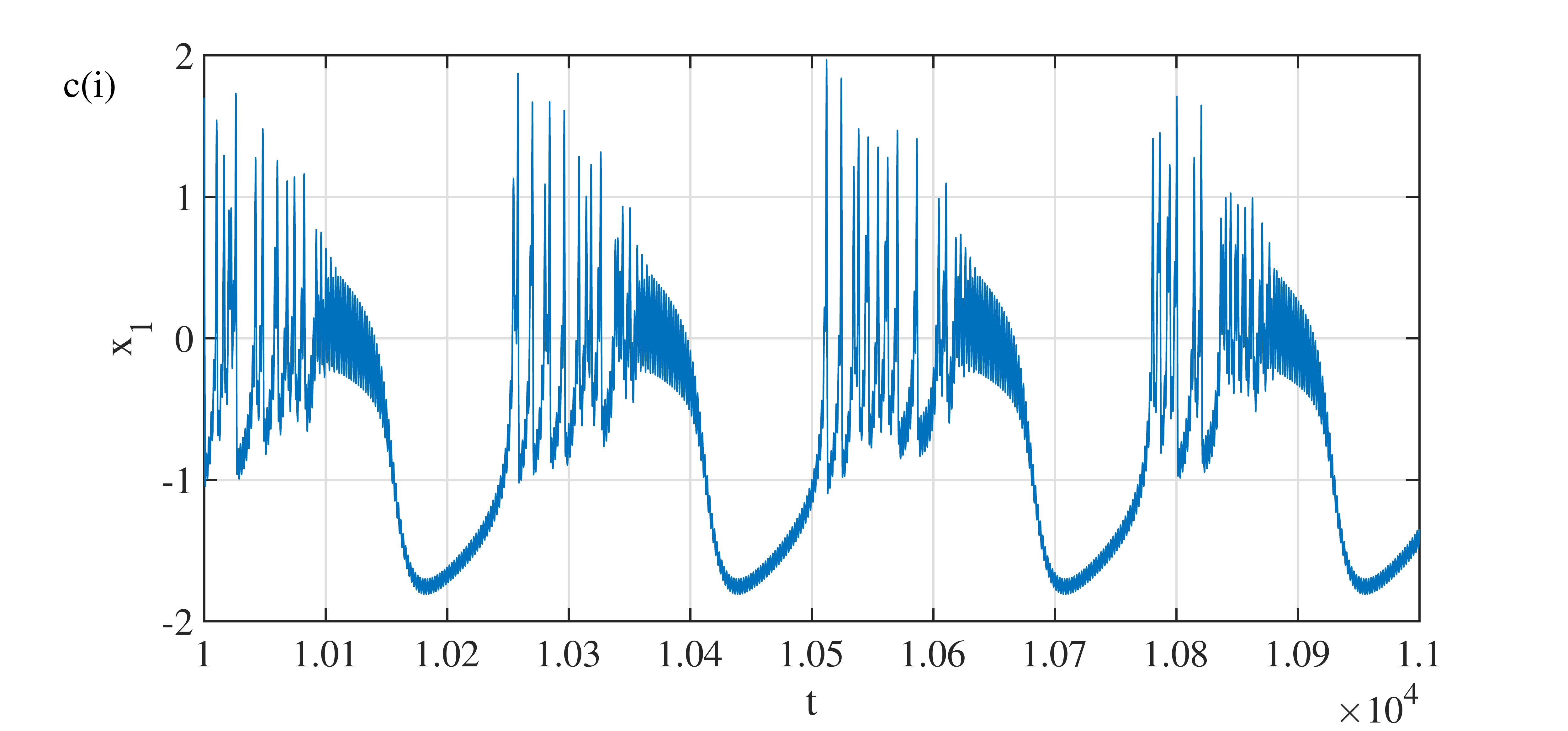}
\caption{}
\end{subfigure}
\begin{subfigure}[b]{0.49\linewidth}
\includegraphics[width=\linewidth]{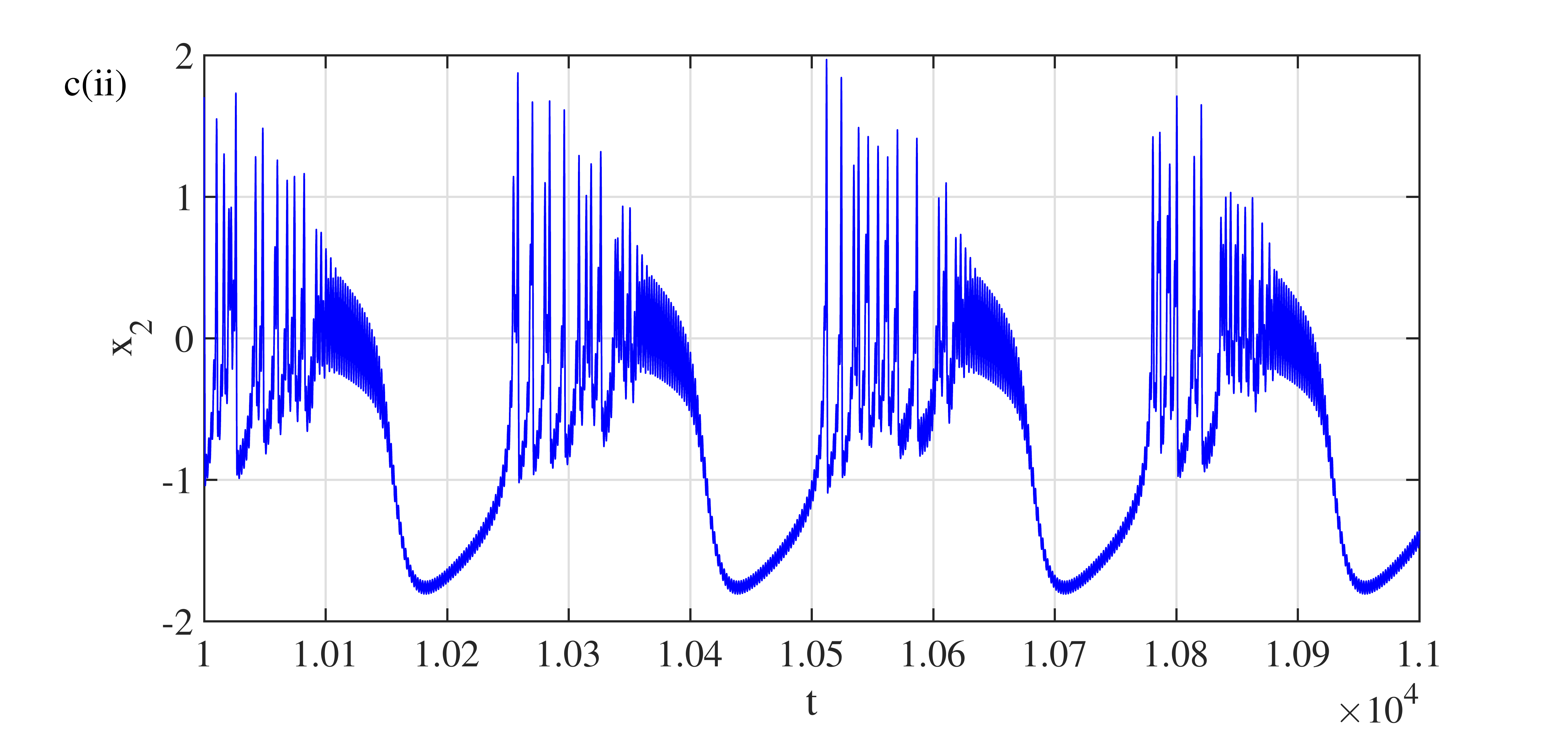}
\caption{}
\end{subfigure}
\medskip
\caption{Time series waveforms of the state variables $x_1$ and $x_2$ of the coupled neurons. For the figures (a) and (b), the parameter value $\sigma = 0$, for (c) and (d), $\sigma = 1$, and for (e) and (f), $\sigma = 100$. The constant parameter values are: $ i_2 = 3$, $m = 2$, and $\alpha = 0.5$. The initial conditions are ($-0.54$,$-5$,$0.1$,$0$,$0$,$6.688$). (Color online)}
\label{fig4} 
\end{figure}
\begin{figure}[tbh]
\centering
\begin{subfigure}[b]{0.49\linewidth}
\includegraphics[width=\linewidth]{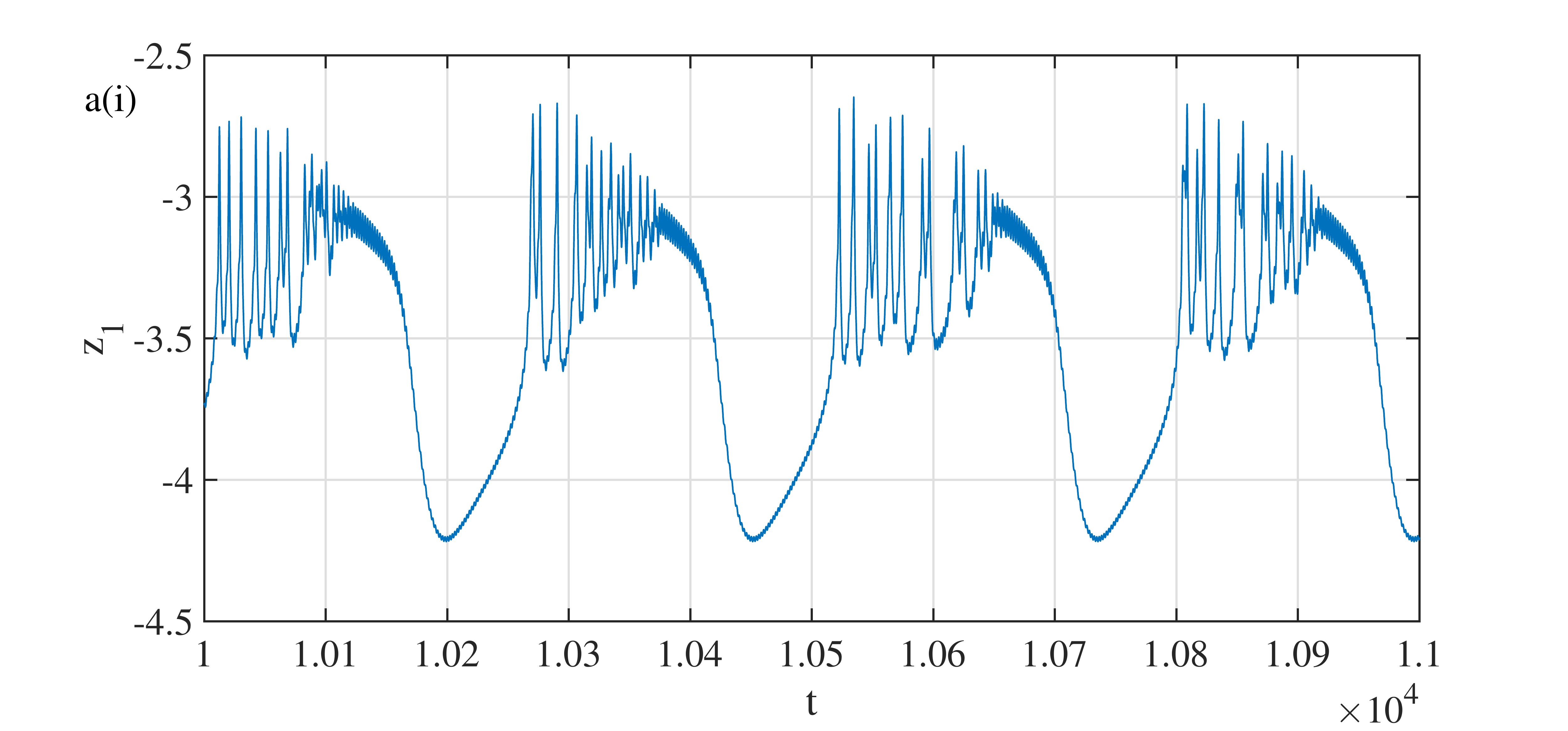}
\caption{}
\end{subfigure}
\begin{subfigure}[b]{0.49\linewidth}
\includegraphics[width=\linewidth]{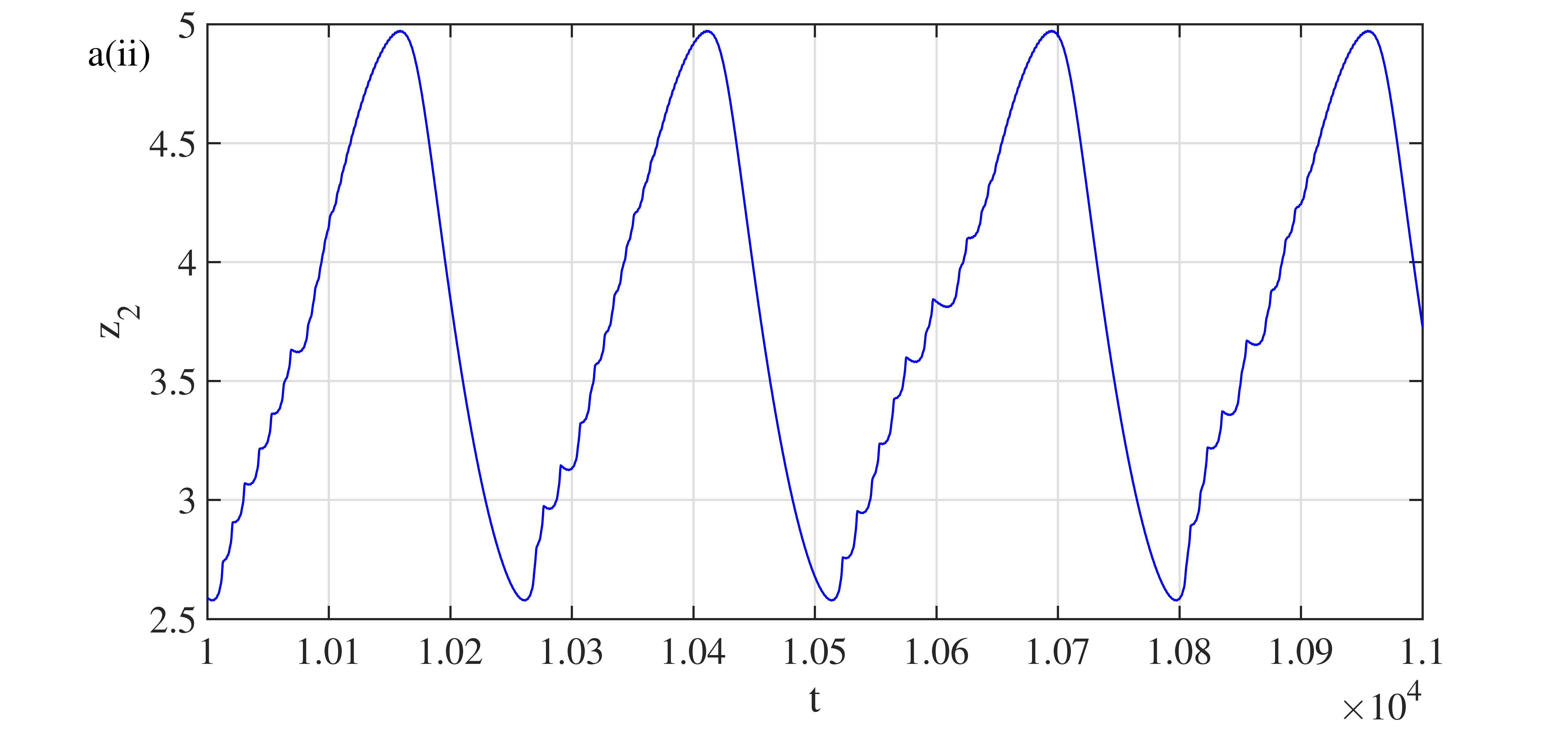}
\caption{}
\end{subfigure}
\medskip
\caption{Time series showing the evolution of the inner variable of the memristive autapse of the first neuron and the slow adaptation current of the second neuron. The parameters are those of the Fig.~\ref{fig4}(e) and \ref{fig4}(f). (Color online)}
\label{fig5} 
\end{figure}
For a deep analysis of the behaviors that can occur in the coupled neurons under the consideration of varying the electrical coupling strength, we have computed the time-series waveforms of $x_1$ and $x_2$, shown in Fig.~\ref{fig4}. These waveforms depict the time evolution of the membrane potential of each coupled neuron for a particular value of the electrical coupling strength, chosen from Fig.~\ref{fig3}(a). Fig.~\ref{fig4}(a) and \ref{fig4}(b) are the time-series waveforms of $x_1$ and $x_2$, when the parameter $\sigma=0$, i.e., when there is no couple between the two neuron models. In this case, the membrane potential of the first neuron exhibits chaotic spiking while the second neuron exhibits chaotic busting. The irregular periodicities in the time series waveforms confirm the chaotic behaviors. When the parameter $\sigma =1$, both the neurons exhibit chaotic bursting with different shapes, as shown in Fig~\ref{fig4}(c) and Fig~\ref{fig4}(d) respectively. The absence of synchronization between the behavior of the coupled neurons is because of the feeble electrical coupling strength $\sigma$. For large enough coupling strength, i.e., when $\sigma = 100$, chaotic burstings of Fig.~\ref{fig4}(e) and Fig.~\ref{fig4}(f) are obtained. The identical shape of the time series supports the partial/cluster of synchronization of the coupled model. Indeed, for $\sigma = 100$, the membrane potential and the recovery potentials of the first and the second neurons are synchronized. In contrast, the inner variable of the memristive autapse of the first neuron and the slow adaptation current of the second neuron is not synchronized. 

Fig.~\ref{fig5} shows the time-series waveforms of $z_1$ and $z_2$ of the coupled neurons. These state variables signify the inner variable of the first neuron's memristive autapse and the slow adaptation current of the second neuron, respectively. As from the figure, we can say that for $\sigma = 100$, these state variables are not in synchronism, which supports the lack of synchronization statements of the heterogeneous HR neuron model.

\begin{figure}[tbh]
\centering
\begin{subfigure}[b]{0.49\linewidth}
\includegraphics[width=\linewidth]{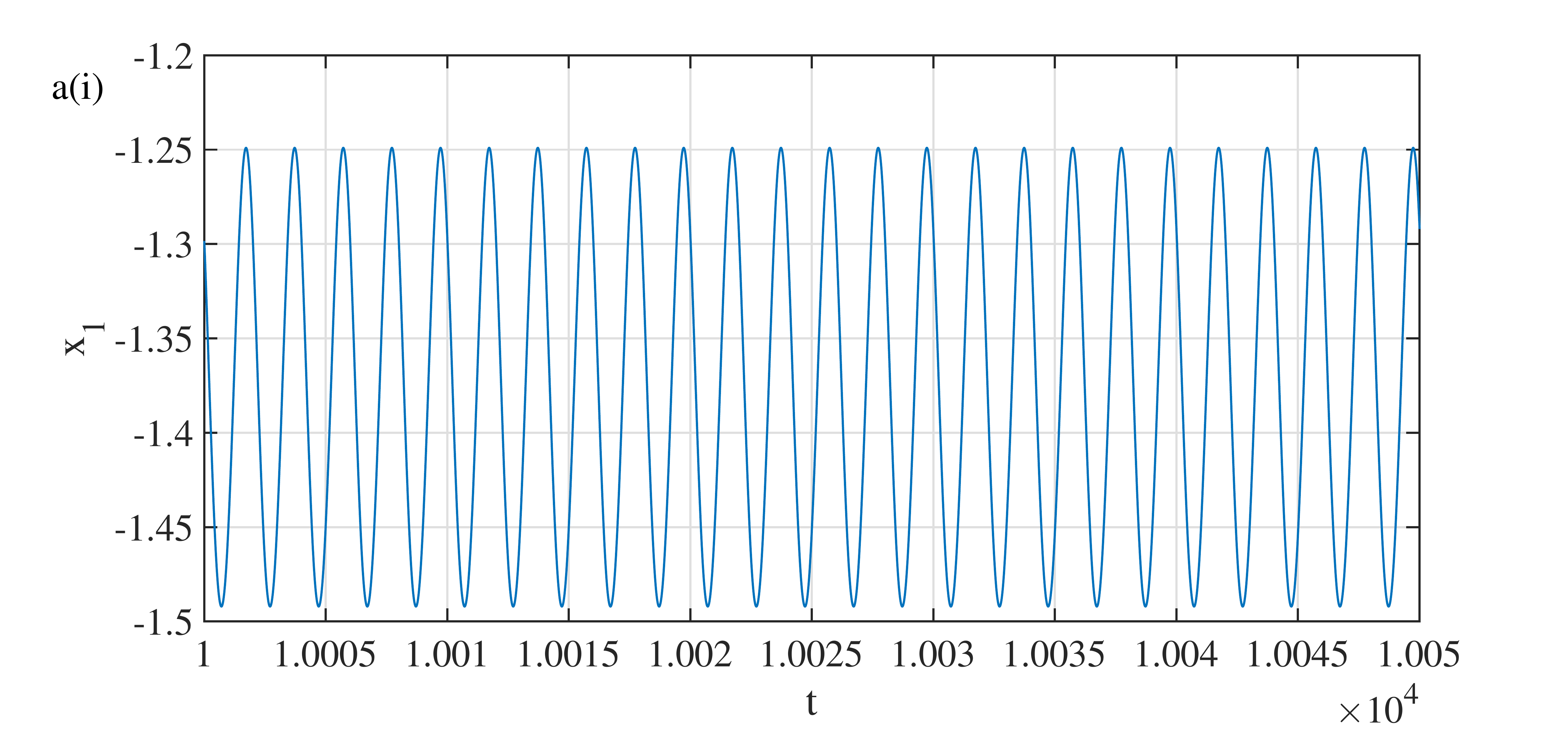}
\caption{}
\end{subfigure}
\begin{subfigure}[b]{0.49\linewidth}
\includegraphics[width=\linewidth]{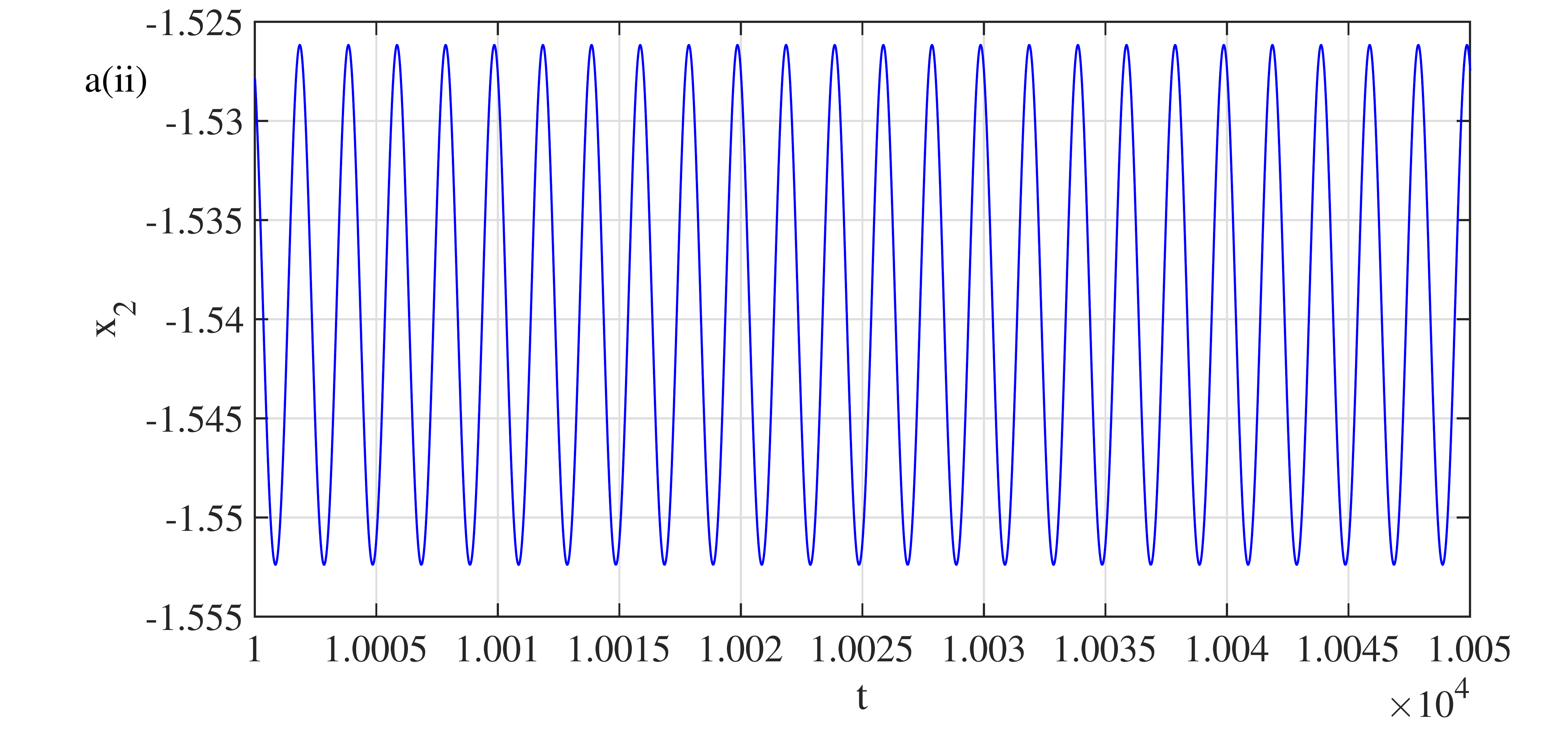}
\caption{}
\end{subfigure}
\begin{subfigure}[b]{0.49\linewidth}
\includegraphics[width=\linewidth]{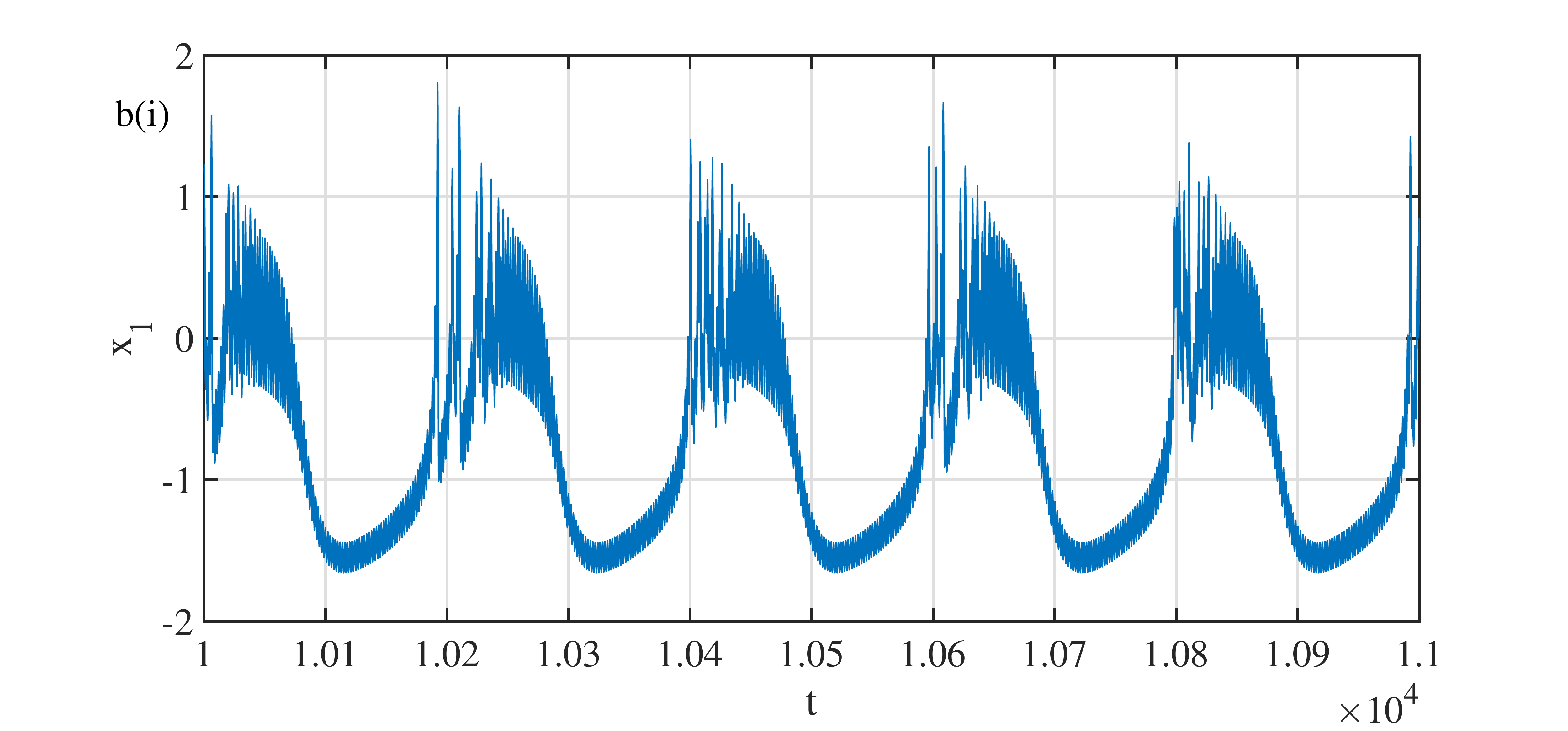}
\caption{}
\end{subfigure}
\begin{subfigure}[b]{0.49\linewidth}
\includegraphics[width=\linewidth]{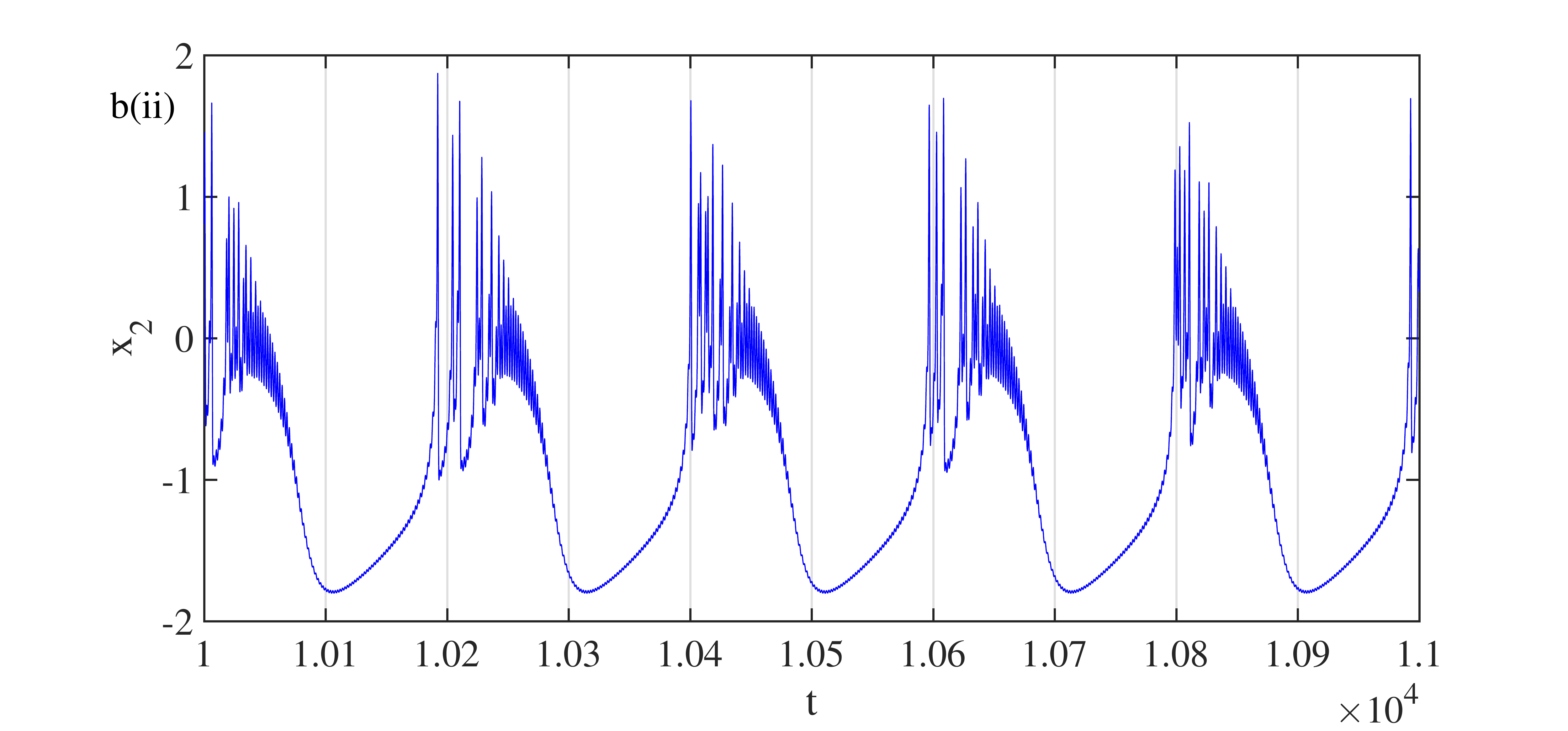}
\caption{}
\end{subfigure}
\begin{subfigure}[b]{0.49\linewidth}
\includegraphics[width=\linewidth]{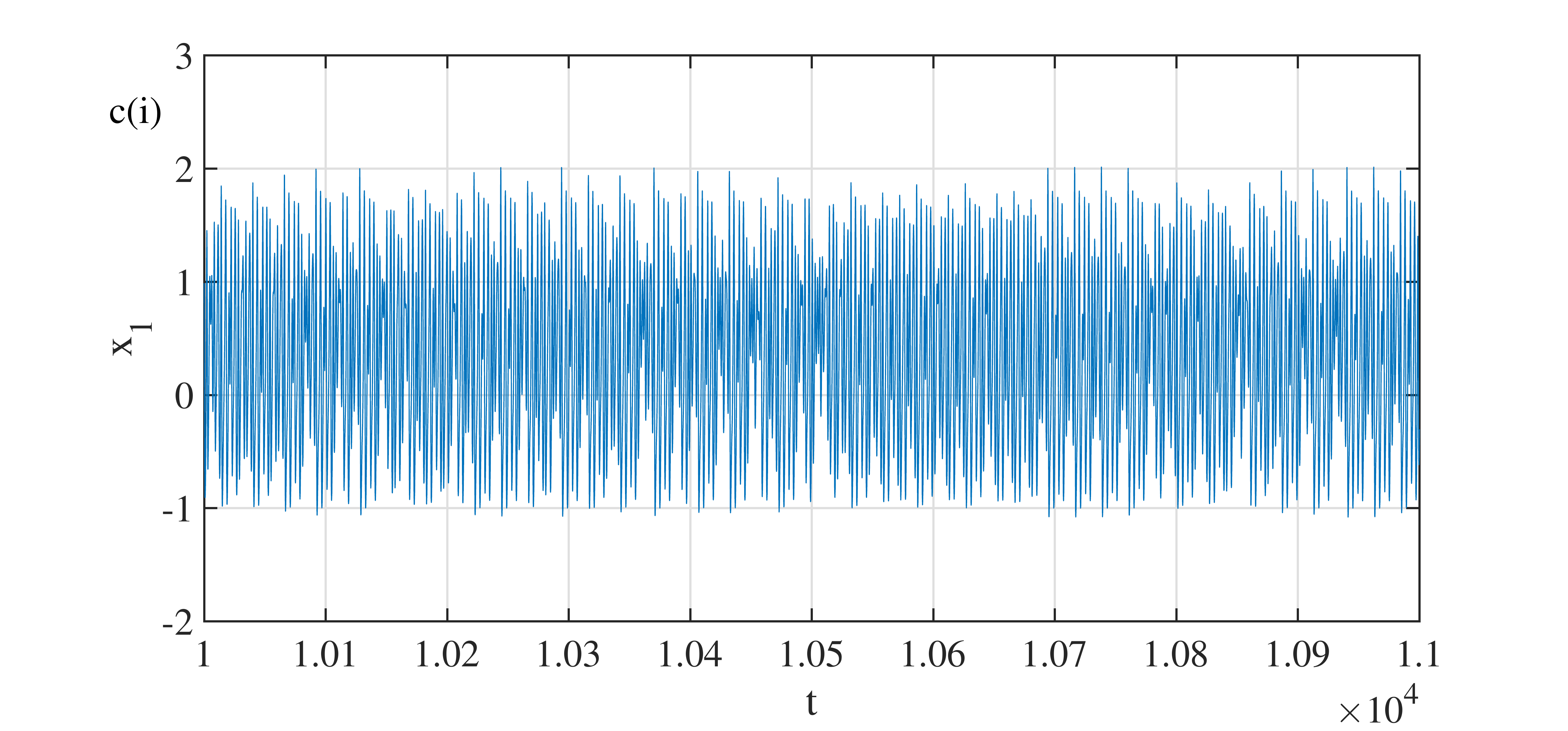}
\caption{}
\end{subfigure}
\begin{subfigure}[b]{0.49\linewidth}
\includegraphics[width=\linewidth]{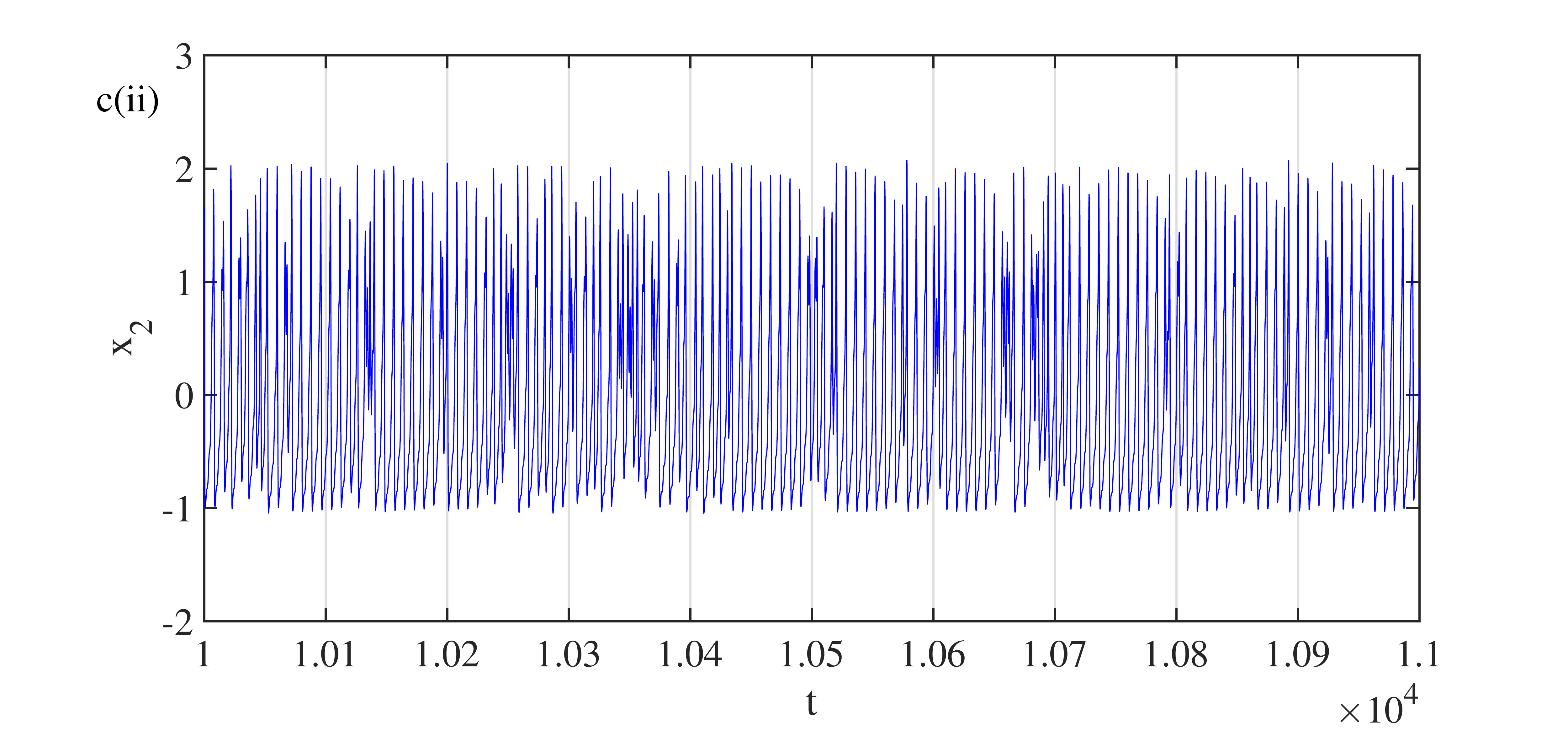}
\caption{}
\end{subfigure}
\medskip
\caption{Time series waveforms of $x_1$ and $x_2$ of the coupled neurons. The constant parameter values are $\sigma = 2$, $m = 2$, and $\alpha = 0.5$. For (a) and (b), $i_2 = 0$, for (c) and (d), $i_2 = 2$, and for (e) and (f), $i_2 = 7$. The initial conditions are ($-0.54$,$-5$,$0.1$,$0$,$0$,$6.688$). (Color online)}
\label{fig6} 
\end{figure}
When $\sigma$ has a lower value close to zero, the two coupled neurons are completely out of synchronization. We now explore the behaviors in the coupled neurons under the absence of synchronization conditions, i.e., for a small constant parameter value of $\sigma$. For that, we have computed the time series of the model using some constant parameter values of the external forcing current, $i_2$. When  $i_2 = 0$, both the neurons exhibit periodic firing activities, as shown in Fig.~\ref{fig6}(a) and \ref{fig6}(b). Chaotic bursting and spiking are exhibited by both neurons of the coupled model as shown in Figs.~\ref{fig6}(c), \ref{fig6}(d), \ref{fig6}(e), and \ref{fig6}(f) for the parameter values $i_2 = 2$ and $i_2 = 7$, respectively. The absence of synchronization between these firing activities is related to the weak electrical coupling strength $\sigma$. Since two coupled neurons are insufficient to study their collective behaviors, the next section of our work, we shall analyze heterogeneous ring-star networks made up of the interconnection of up to $100$ coupled neurons under various coupling configurations.

\section{Network analysis}
\label{na}
\begin{figure}[tbh]
\centering
\begin{subfigure}[b]{0.49\linewidth}
\includegraphics[width=\linewidth]{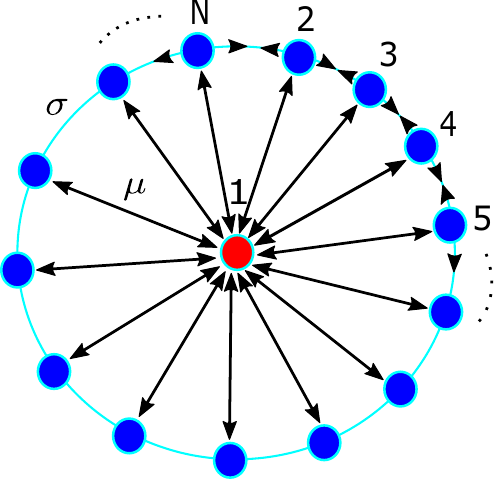}
\caption{}
\end{subfigure}
\begin{subfigure}[b]{0.49\linewidth}
\includegraphics[width=\linewidth]{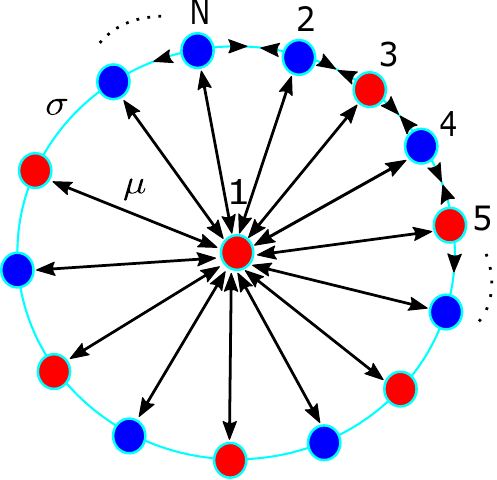}
\caption{}
\end{subfigure}
\begin{subfigure}[b]{0.51\linewidth}
\includegraphics[width=\linewidth]{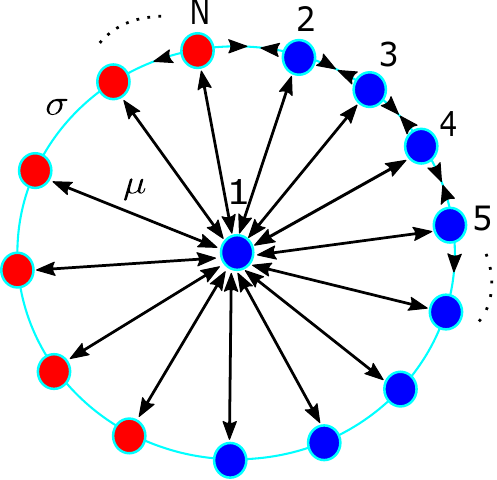}
\caption{}
\end{subfigure}
\medskip
\caption{(a) Heterogeneous ring-star network of $N=100$ nodes with the central node in red color as the traditional HR neuron and the other end nodes set in blue color as the memristive HR neuron, (b) Ring-star configuration of network~B is considered. The even-numbered network nodes in blue color are the memristive HR neuron model. The odd-numbered nodes of the network are composed of the traditional HR neuron model, which is portrayed in red color, (c) Ring-star configuration of network~C is considered. Each node on the right hemisphere in blue color constitutes the memristive HR neuron nodes, and on the left in red forms the traditional HR neuron. (Color online)}
\label{ref:Networks} 
\end{figure}
In this section, it is good to emphasize that, in the connection topology of the Fig.~\ref{ref:Networks}(a), Fig.~\ref{ref:Networks}(b), and Fig.~\ref{ref:Networks}(c), the red nodes are associated with the $3$-D traditional HR neuron; and the blue nodes are associated with the memristive $2$-D HR neuron. After exploring a single coupled heterogeneous neuron model in the section~\ref{db}, we have considered in this section a heterogeneous ring-star system of both the traditional Hindmarsh neuron and memristive Hindmarsh neuron models.

The dynamical equations of the ring-star network are given by
\begin{align}
	\begin{split}
		\label{eq:diffusive}
		\dot{x_{i}} &= f_{x} + \mu(x_{i}-x_{1}) + \frac{\sigma}{2P} \sum_{n=i-P}^{n=i+P}(x_{i}-x_{n}),\\
		\dot{y_{i}} &= f_{y},\\
		\dot{z_{i}} &= f_{z}.
	\end{split}
\end{align}
\noindent For $i=1$ (central node) the dynamical equations are: 
\begin{align}
	\begin{split}
		\label{eq:central}
		\dot{x_{1}} &= f_{x} + \sum_{j=1}^{N} \mu(x_{j} - x_{1}),\\
		\dot{y_{1}} &= f_{y},\\
		\dot{z_{1}} &= f_{z}.
	\end{split}
\end{align}
\noindent where depending upon one of the three network configurations, the functional form of $f$ can be either traditional HR neuron or memristive HR neuron, 
\noindent with periodic boundary conditions:
\begin{align*} 
	x_{i+N}(t) &= x_{i}(t),\\
	y_{i+N}(t) &= y_{i}(t),\\
	z_{i+N}(t) &= z_{i}(t),
\end{align*}
\noindent for $i=2,3,\ldots,N$.

The network size is considered to be of $100$ nodes with $P$ nearest neighbors connected with each other having periodic boundary conditions. The network parameters, such as the ring coupling strength $\sigma$, star coupling strength $\mu$, and the coupling range $P$, will be varied to explore different synchronization patterns arising in the ring-star network memristive Hindmarsh-Rose neuron system.

\begin{figure*}[tbh]
\centering
\includegraphics[width=0.7\textwidth]{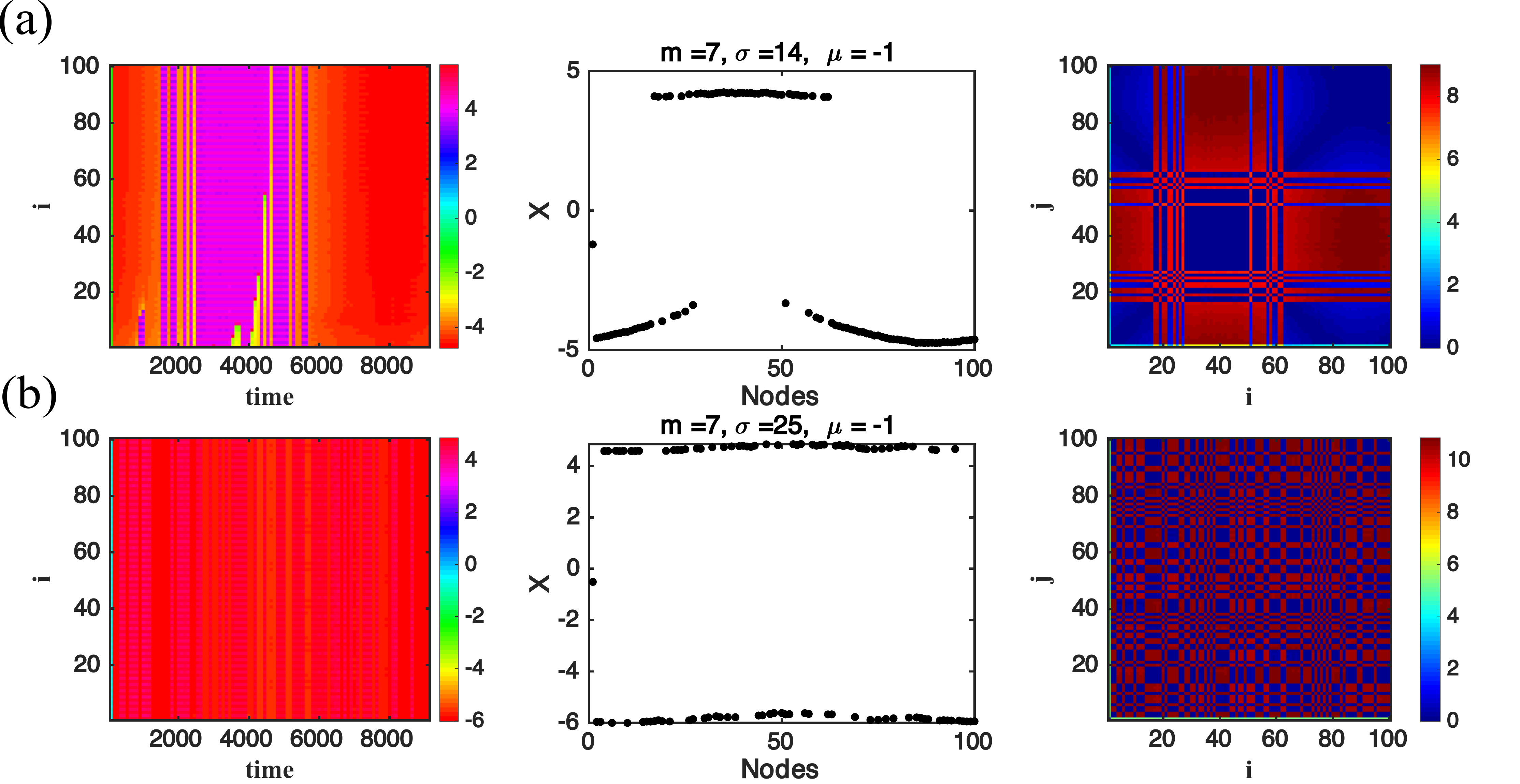}
\caption{Double-well chimera state in (a) and two cluster state in (b) in ring-star network A. The spatiotemporal plots on the left, nodes plot on middle, and the recurrence plot on the right. (Color online)}
\label{fig:NetARingStar}
\end{figure*}
\begin{figure*}[tbh]
\centering
\includegraphics[width=0.7\textwidth]{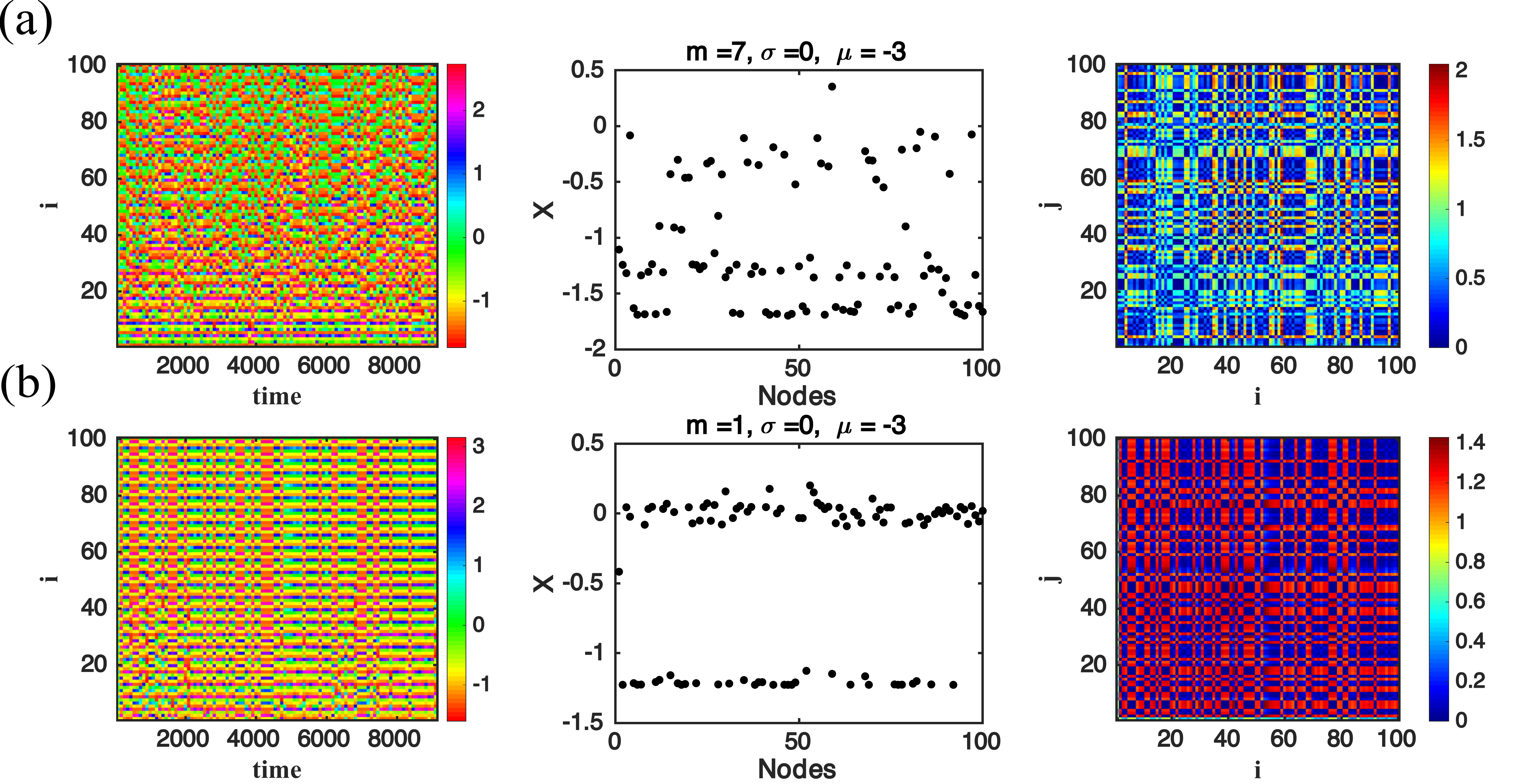}
\caption{Star network configuration of network A. In a) the oscillators are unsynchronized when $  m=7 $ b) the oscillators synchronize themselves in a two cluster state when $  m=1 $. The star coupling strength is fixed to be $\mu = -3$. (Color online)}
\label{fig:NetAStar}
\end{figure*}
\begin{figure*}[tbh]
\centering
\includegraphics[width=0.7\textwidth]{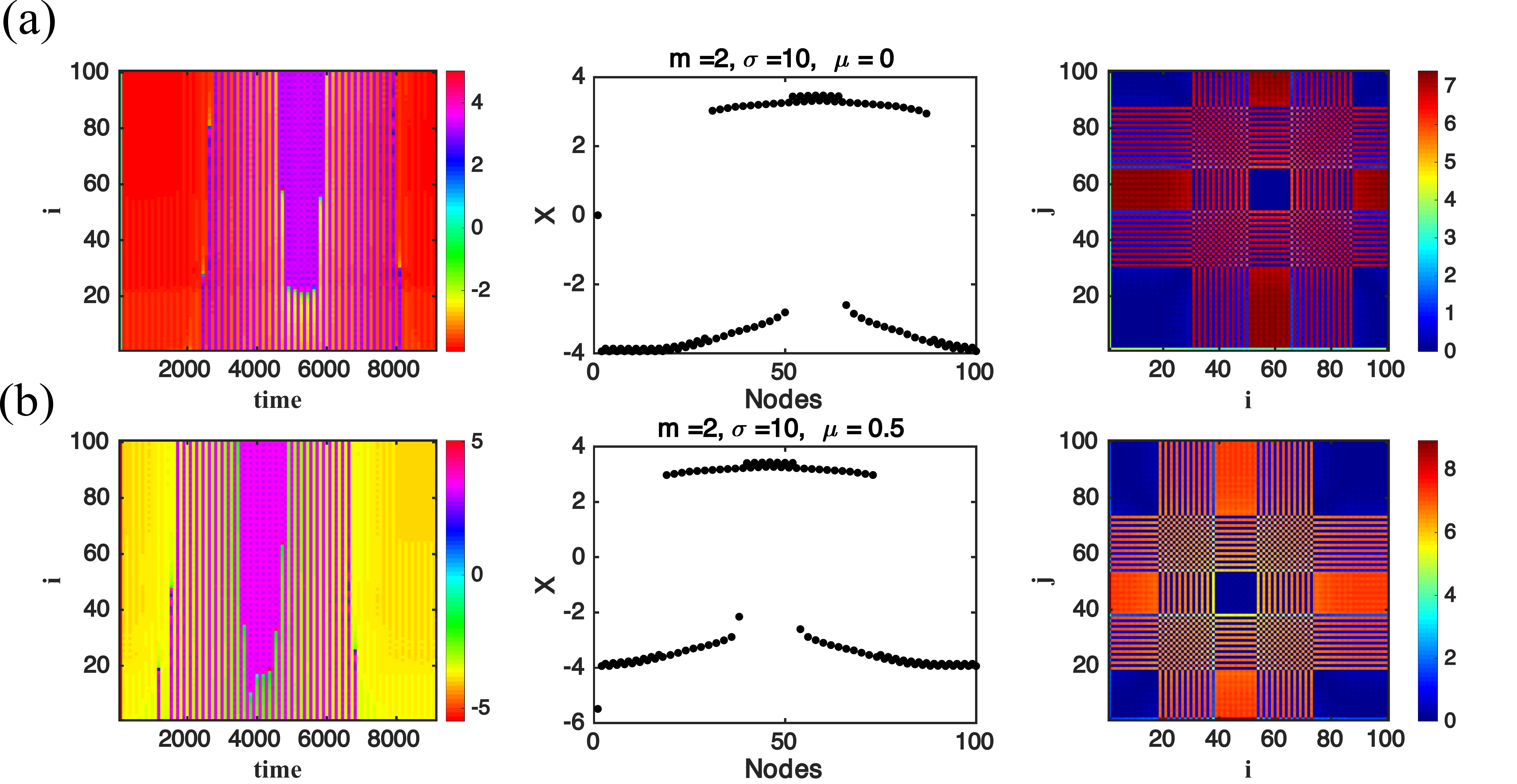}
\caption{Ring network configuration in (a) and ring-star configuration in (b) are considered for network~B. The presence of clusters and double-well chimera state can be seen in both of the network configurations. (Color online)}
\label{fig:NetBRingStar}
\end{figure*}

In this study, we show three different heterogeneous networks. The motivation is to introduce heterogeneity in the network gradually. Network~A is composed of a traditional HR neuron as the central node, and the other end nodes are comprised of memristive HR neurons. The schematic representation is shown in Fig.~\ref{ref:Networks}(a). Network~B is constructed with the odd number of nodes made of traditional HR neurons and the even number of nodes made of the memristive HR neuron models, as shown in Fig.~\ref{ref:Networks}(b). Network~C has been set up for $N=100$~nodes, the right hemisphere is connected to memristive HR neurons, and the left hemisphere comprises traditional HR neurons. The schametic representation of the Network~C is shown in Fig.~\ref{ref:Networks}(c). Since the heterogeneity is due to the presence of both traditional HR neurons and memristive HR neurons, the functional form of $f$ for the $i$-th node in the equations~\eqref{eq:diffusive} and \eqref{eq:central} changes according to the configuration of the network A, B, C. The functional form of $f$ for $i$-th node can be expressed as either in the equation~\eqref{eq1} or the equation~\eqref{eq2}. The working parameter set for this section of the paper are fixed as follows: for the traditional HR neuron in equation~\eqref{eq1}, the parameters are $b_{2} = 3, d_{2} = 5, c_{2}=1, a_{2}=1, \gamma=0.008, \bar{x} = -1.6, i_{2} = 5.5, s=4$. For the memristive HR neuron, the parameters are $a_{1} = 1, c_{1}=5, d_{1}=5, e=0.5, m=0.02, f=0.5, \alpha = 0.01$.

\noindent\textbf{Network A:} We consider a ring-star network of the system as shown in Fig.~\ref{ref:Networks}(a). We reveal different spatiotemporal patterns upon the variation of the star and ring coupling strengths $\sigma$.

In Fig.~\ref{fig:NetARingStar}(a), we have shown a double-well chimera state with the coexistence of coherence and incoherent nodes. The nodes oscillate in both the negative and positive values of the state variables $x$  in the ring-star network when the ring coupling strength $\sigma$ and star coupling strength $\mu$ are $14$ and $-1$, respectively. It is confirmed via spatiotemporal plot on the left and recurrence plot on the right comprising blue (coherent) and red (incoherent) colors confirming a chimera. Since the structure traverses both the positive and negative values (both wells), it is called a double-well chimera state. When the ring coupling strength is increased, the chimera state gets destroyed and settles down to a two-cluster state, as shown in Fig.~\ref{fig:NetARingStar}(b). It is also confirmed via the recurrence plot on the right due to many small square structures.

In Fig.~\ref{fig:NetAStar}(a), a star network of Fig.~\ref{ref:Networks}(a) is considered by setting the ring coupling strength $\sigma$ at $0$. When $\mu$ = $-3$,  the nodes oscillate incoherently as evident from the nodes plot and the spatiotemporal pattern plots in Fig.~\ref{fig:NetAStar}(a). When the memristive strength $m$ is decreased to $1$, the oscillators start to gather around in a two-cluster state as shown in Fig.~\ref{fig:NetAStar}(b). It can also be confirmed via the spatiotemporal plot and square-shaped repetitive structures in the recurrence plot.
\begin{figure*}[tbh]
\centering
\includegraphics[width=0.7\textwidth]{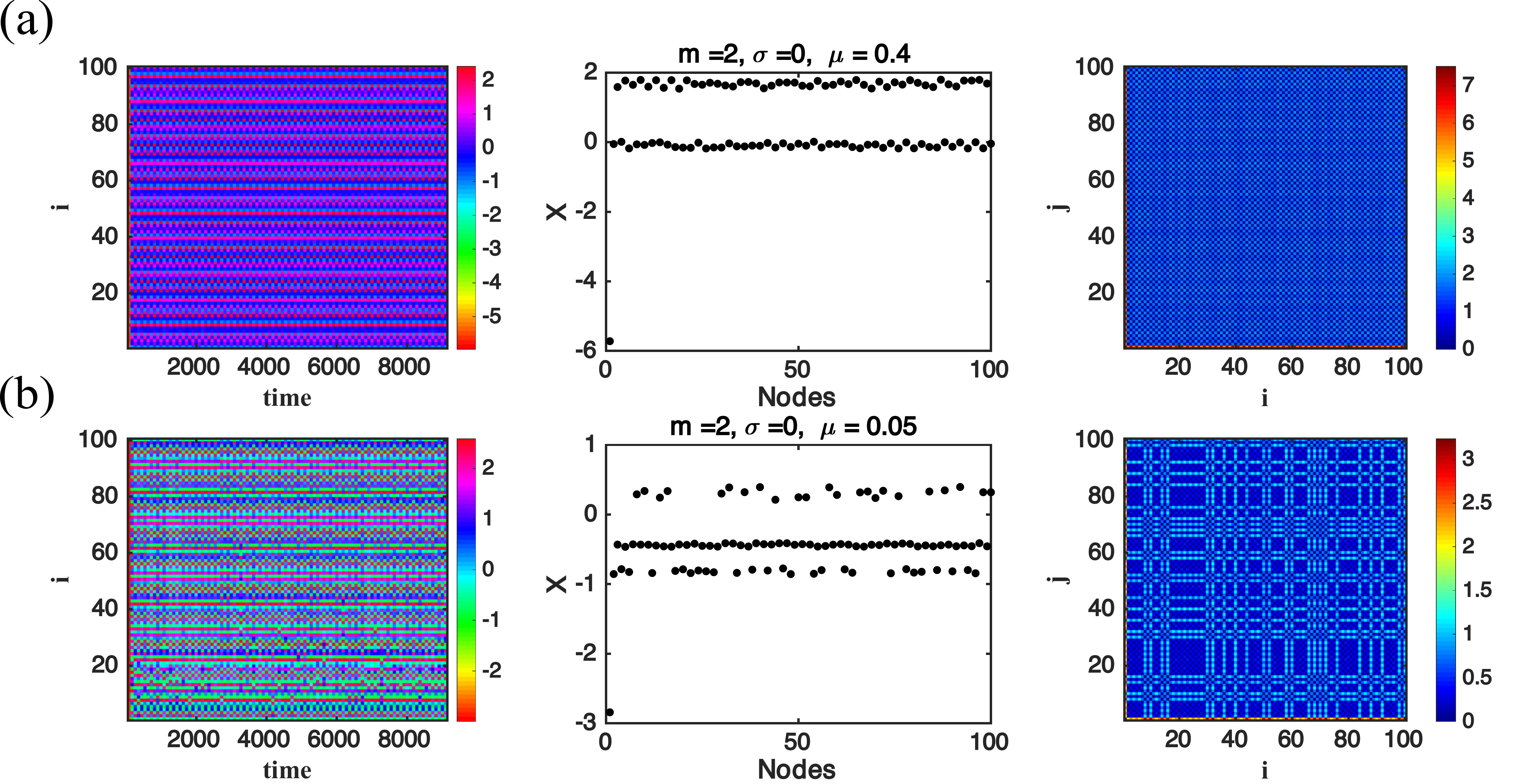}
\caption{Star network configuration of network B is considered. When $\mu=0.4$, we observe a two cluster state in (a) and when $\mu=0.05$, we observe a three cluster state in (b). (Color online)}
\label{fig:NetBStar}
\end{figure*}

\noindent \textbf{Network B:} We further add more heterogeneity to the network~A. One of the ways to proceed is to consider even-numbered nodes to be of memristive HR neurons and the odd-numbered nodes to be of standard HR neurons. The schametic representation of this kind of model is shown in Fig. \ref{ref:Networks}(b). It can be observed that the number of memristive HR neuron nodes has decreased in this case as compared to network~A. We then explore different spatiotemporal patterns in this case.

For the ring network type B, we have observed a double-well-like state with clusters as shown in Fig.~\ref{fig:NetBRingStar}. Although the system steers the double-well-shaped state, it is qualitatively different from a double-well chimera state (as compared with Fig.~\ref{fig:NetARingStar}). The recurrence plots on the right confirm the previous statement. The new type of double-well chimera state can be called a double-well clustered chimera state.

We can verify the cluster property by forming the small square structures using the recurrence plot of network type B. In the case of the ring network, such double-well chimera state is shown in Fig. \ref{fig:NetBRingStar}(a) for the parameter value $\sigma=10$ and in the case of ring-star network, the double-well chimera state is shown in Fig. \ref{fig:NetBRingStar}(b). We have noted that the addition of heterogeneous nodes leads to the formation of new types of chimera-like states, such as the prevalence of double-well clustered chimera state revealed in this network configuration.

We next consider the star configuration of network~B by setting $\sigma$ at $0$, and $\mu \neq 0$, as shown in Fig.~\ref{fig:NetBStar}. When the star coupling strength, $\mu$, is set to $0.4$, we observe a double cluster state in Fig.~\ref{fig:NetBStar}(a). It is evident from the space-time plot on the left and the recurrence plot on the right via the formation of tiny squares of Fig~\ref{fig:NetBStar}(a). The cluster state configuration is robust with the coupling strength $\mu$ variation. When the coupling strength is decreased to $\mu =0.05$, we observe a three clustered state, which is shown in Fig. \ref{fig:NetBStar}(b). The recurrence plot also confirms this three-cluster state by forming many tiny squared structures.

\begin{figure*}[tbh]
\begin{center}
\includegraphics[width=0.7\textwidth]{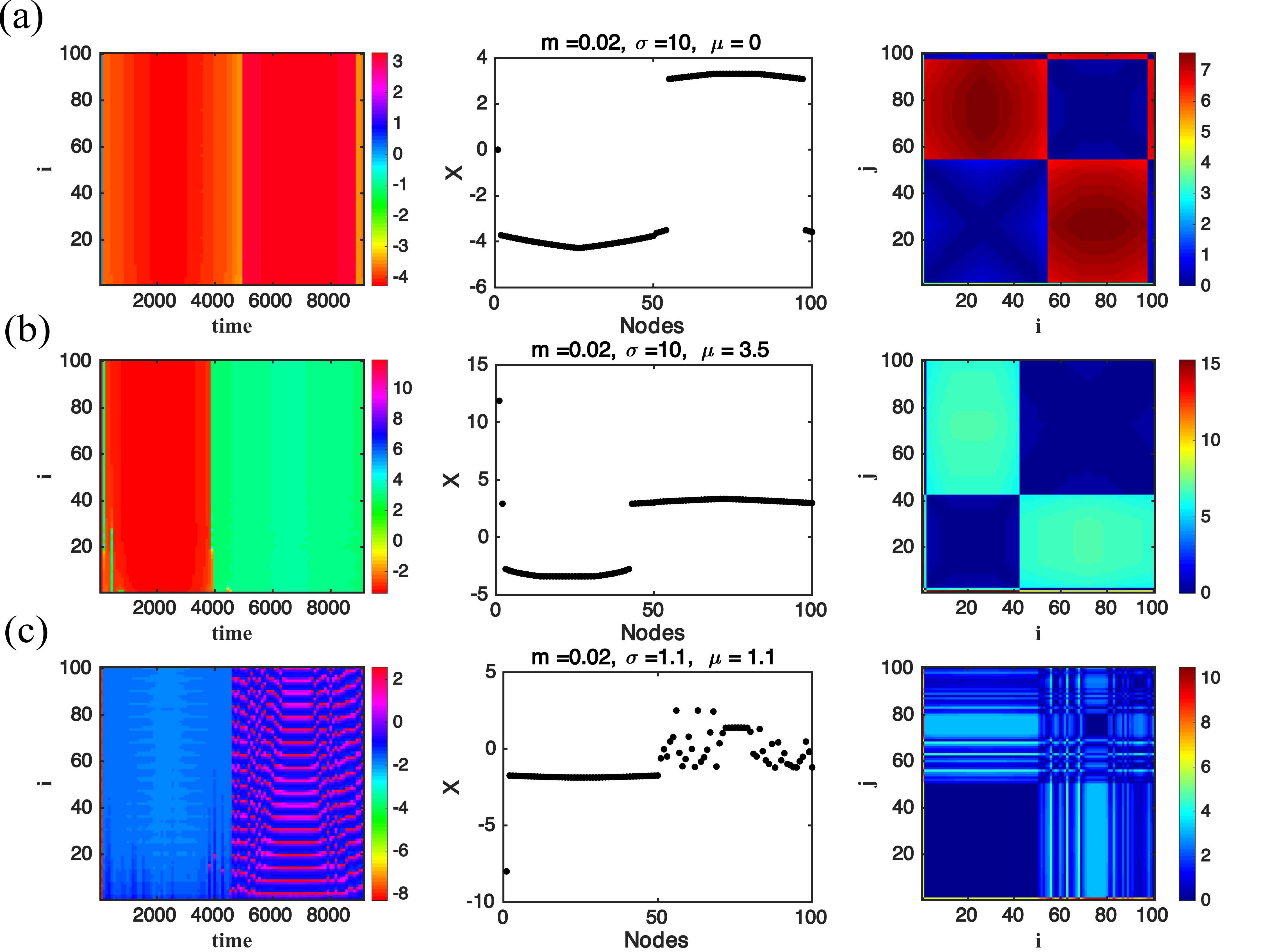}
\end{center}
\caption{A ring-star configuration of network C is considered. In (a), (b) a double-well cluster state is shown. observe that the oscillators are in both positive and negative $x$ values. In (c), a chimera state is shown evident from the recurrence and spatiotemporal plot. }
\label{fig:NetCRingStar}
\end{figure*}
\noindent \textbf{Network C:} In this network, we consider a bipartite network, where the ring star network is divided into two hemispheres. Memristive HR neurons are connected on the right hemisphere (in blue), and on the left, traditional HR neurons are interconnected (in red color). The schematic representation of this type of network is shown in Fig.~\ref{ref:Networks}(c). We shall show that this network configuration resulted in the prevalence of chimera states and cluster nodes.

\begin{figure}[tbh]
\centering
\includegraphics[width=0.3\textwidth]{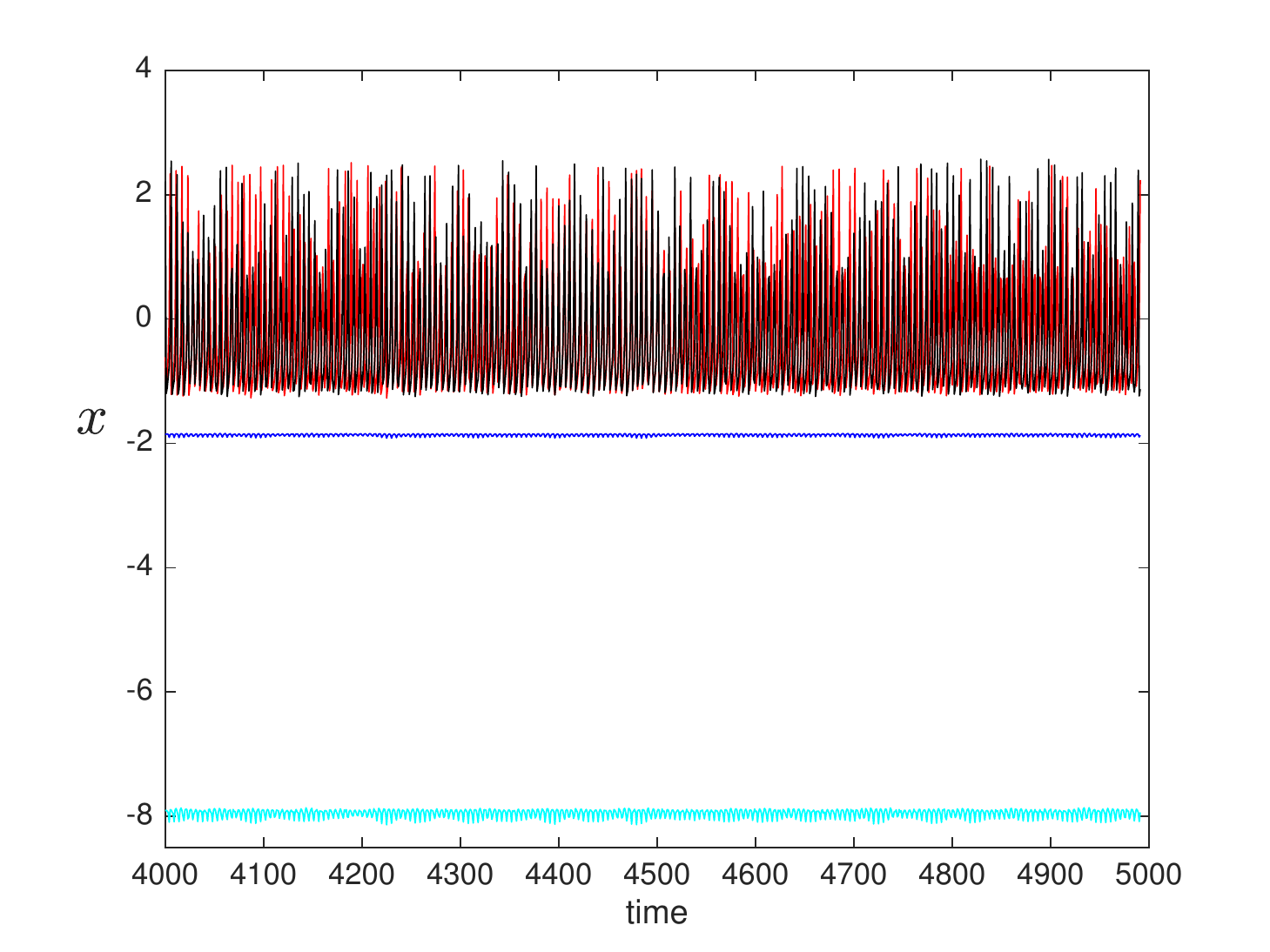}
\caption{Time series of chimera state for the ring-star network configuration C. Four nodes $i=1$ (central node) marked in blue, $i=25$ node marked in blue, $i=55$ node marked in red, $i=95$ node marked in black. It clearly shows the coexistence of synchronous and asynchronous states o a chimera. (Color online)}
\label{fig:TimeSeriesNetC}
\end{figure}
\begin{figure*}[tbh]
\centering
\includegraphics[width=0.7\textwidth]{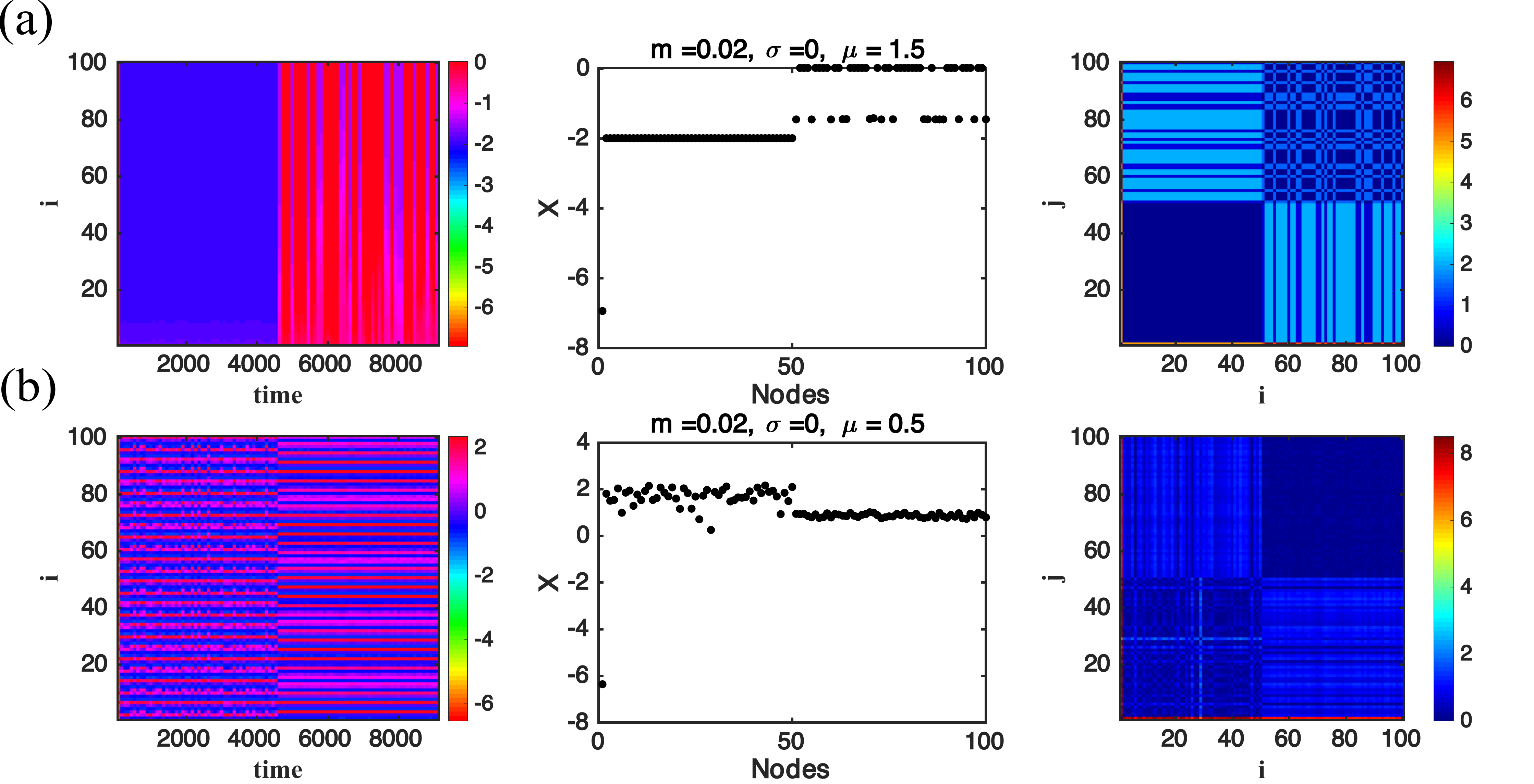}
\caption{Star network configuration of network C is considered. In (a), a three cluster state is shown when $\mu =1.5$. In (b), a two cluster state is seen when $\mu$ is decreased to $0.5$. (Color online)}
\label{fig:NetCStar}
\end{figure*}
In the case of the ring network ($\mu =0$), we observe that a double-well cluster state comes to an exist, that is evident from the state variable plots and also from the recurrence plot, as shown in Fig.~\ref{fig:NetCRingStar}(a). A two cluster state is observed in a ring-star configuration when $\sigma=10$, $\mu=3.5$ in  Fig.~\ref{fig:NetCRingStar}(b). When $\sigma =1.1, \mu  =1.1$, the prevalence of chimera state is recorded and can be verified by observing the state variable plot of the nodes of the network and also from the recurrence plot on the right in  Fig.~\ref{fig:NetCRingStar}(c).

It is essential to note that in the case of bipartite network fosters, the prevalence of chimera states and cluster states with an increase in the number of clusters are formed, which is shown in Fig.~\ref{fig:NetCRingStar}(c). A time series plot for Fig.~\ref{fig:NetCRingStar}(c) supports the existence of a chimera state, where both synchronous and asynchronous regimes having four nodes were identified from Fig.~\ref{fig:NetCRingStar}(c). The evolution of the central node is marked in cyan, the node number less than $50$ in sync is shown in blue, node number $55$ kept in red, node number $95$ marked in black. The time series is shown in Fig.~\ref{fig:TimeSeriesNetC}, where we can see the coexistence of both synchronous and asynchronous nodes confirming a chimera state.

In the case of a star network, when $\mu  =1.5$, we observe a three-cluster state, as shown in Fig.~\ref{fig:NetCStar}(a) making a transition in both positive and negative values of $X$. It is evident from both the state variable plot and the spatiotemporal plot. When $\mu$ is decreased to $0.5$, we observe a close two cluster states as in Fig.~\ref{fig:NetCStar}(b).

\section{Experimental Results}
\label{expt_results}

\begin{figure*}[tbh]
\centering
\includegraphics[width=0.9\linewidth]{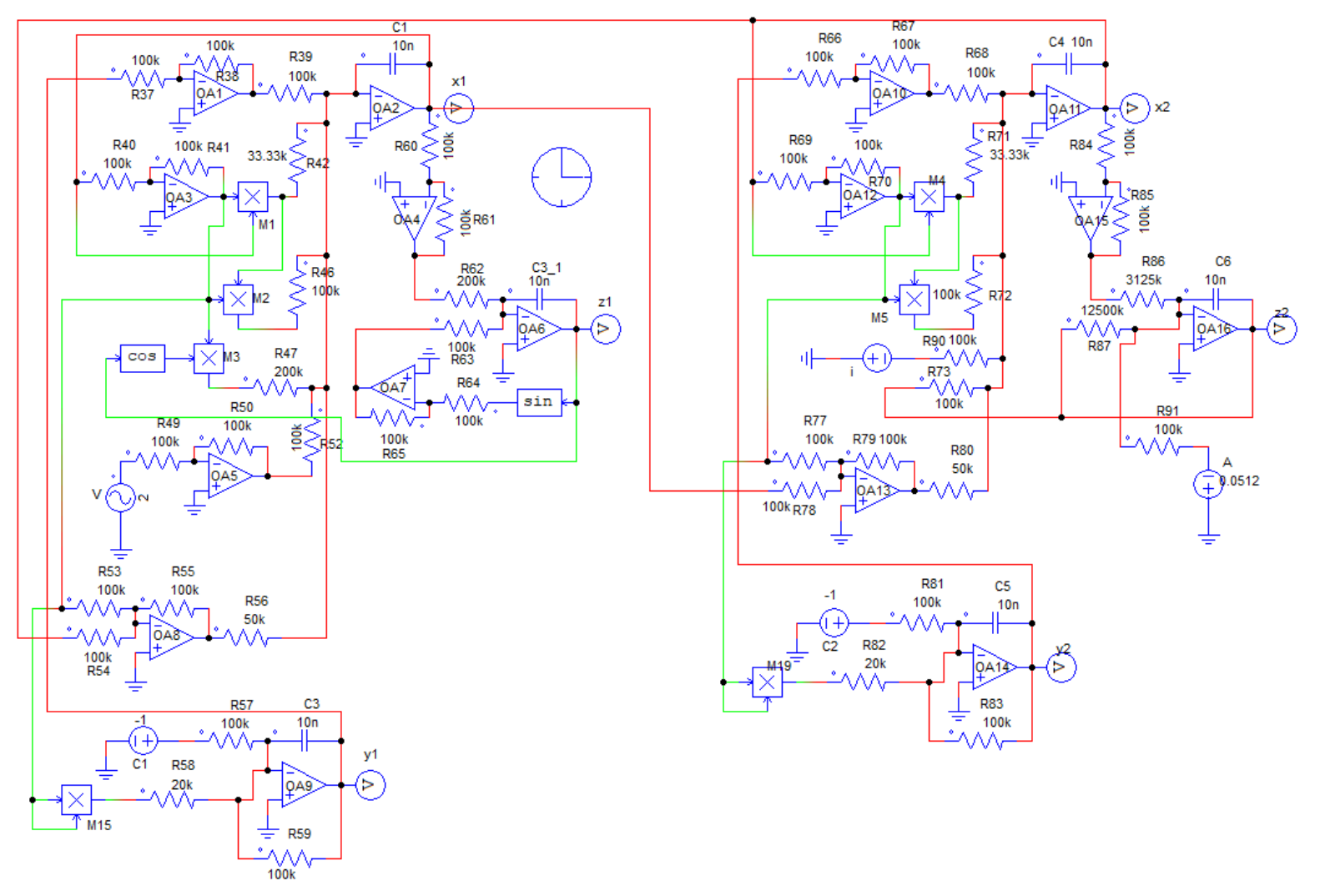}
\caption{Circuit implementation of the considered system. OA$1$ to OA$16$ are the Op-Amps, M$1$ to M$5$ are the multipliers. The values of resistances are mentioned in the circuit diagram. The capacitors $C_1$, $C_2$, $C_3$, $C_4$, $C_5$, and $C_6$ are $10$~nF each. The AC sine wave $V$ has an amplitude $2$~V and the frequency $500$~Hz. The supply voltages have been taken as $+V_{\rm cc} = +15.0$~V and $-V_{\rm cc} = -15.0$~V. The initial condition is chosen as the initial values of the six capacitor voltages in V, $(-0.54,-5,0.1,0,0,6.688)$. (Color online)}
\label{fig:Fig 1}
\end{figure*}
In order to validate the numerically predicted results, an equivalent circuit diagram of the considered system (as shown in equation.~(\ref{eq3})) is developed. The electronic circuit diagram is shown in Fig.~\ref{fig:Fig 1}. To achieve the circuit, we have rewritten the equation~(\ref{eq3}) as below:
\begin{equation}
\begin{split}
\label{particular-a}
CR\dot{x}_1 = &-\bigg[\Big(\frac{R}{R_{\rm 39}}\Big)\cdot(- y_1) +\Big(\frac{R}{R_{\rm 42}}\Big)\cdot(-x_1)\cdot(x_1) + \Big(\frac{R}{R_{\rm 46}}\Big) \\& \cdot(-x_1^2)\cdot(-x_1) + \Big(\frac{R}{R_{\rm 47}}\Big)\cdot(-x_1)\cdot(\cos(z_1)) +  \Big(\frac{R}{R_{\rm 52}}\Big)\cdot \\& \big(-m\cdot \sin(2\pi ft)\big) + \Big(\frac{R}{R_{\rm 56}}\Big)\cdot(x_1 - x_2) \bigg]
\end{split}
\end{equation}
\begin{equation}
\begin{split}
\label{particular-b}
CR\dot{y}_1 = & -\bigg[\Big(\frac{R}{R_{\rm 57}}\Big)\cdot(-C_1) + \Big(\frac{R}{R_{\rm 58}}\Big)\cdot(-x_1)(-x_1) \\& + \Big(\frac{R}{R_{\rm 59}}\Big)\cdot(y_1) \bigg]
\end{split}
\end{equation}
\begin{equation}
\begin{split}
\label{particular-c}
CR\dot{z}_1 = & -\bigg[\Big(\frac{R}{R_{\rm 62}}\Big)\cdot(-x_1) + \Big(\frac{R}{R_{\rm 63}}\Big)\cdot(-\sin(z_1)) \bigg]
\end{split}
\end{equation}
\begin{equation}
\begin{split}
\label{particular-d}
CR\dot{x}_2 = &-\bigg[\Big(\frac{R}{R_{\rm 68}}\Big)\cdot(- y_2) +\Big(\frac{R}{R_{\rm 71}}\Big)\cdot(-x_2)\cdot(x_2) + \Big(\frac{R}{R_{\rm 72}}\Big)\cdot \\& (-x_2^2)\cdot(-x_2) + \Big(\frac{R}{R_{\rm 73}}\Big)\cdot(z_2) + \Big(\frac{R}{R_{\rm 80}}\Big)\cdot(-x_1 + x_2) \\& +  \Big(\frac{R}{R_{\rm 72}}\Big)\cdot(-i) \bigg]
\end{split}
\end{equation}
\begin{equation}
\begin{split}
\label{particular-e}
CR\dot{y}_2 = & -\bigg[\Big(\frac{R}{R_{\rm 68}}\Big)\cdot(-C_2) + \Big(\frac{R}{R_{\rm 82}}\Big)\cdot(-x_2)(-x_2) \\& + \Big(\frac{R}{R_{\rm 83}}\Big)\cdot(y_2) \bigg]
\end{split}
\end{equation}
\begin{equation}
\begin{split}
\label{particular-f}
CR\dot{z}_2 = & -\bigg[\Big(\frac{R}{R_{\rm 86}}\Big)\cdot(-x_2) + \Big(\frac{R}{R_{\rm 87}}\Big)\cdot(z_2) + \Big(\frac{R}{R_{\rm 91}}\Big)\cdot(-A) \bigg]
\end{split}
\end{equation}
where, $\frac{R}{R_{\rm 42}}$ = $b_1$, $\frac{R}{R_{\rm 71}}$ = $b_2$, $\frac{R}{R_{\rm 46}}$ = $a_1$, $\frac{R}{R_{\rm 72}}$ = $a_2$, $\frac{R}{R_{\rm 47}}$ = $\alpha$, $\frac{R}{R_{\rm 52}}$ = $\frac{R}{R_{\rm 80}}$ = $\sigma$, $\frac{R}{R_{\rm 58}}$ = $d_1$, $\frac{R}{R_{\rm 82}}$ = $d_2$, $\frac{R}{R_{\rm 62}}$ = $e$, $\frac{R}{R_{\rm 86}}$ = $rs$, $\frac{R}{R_{\rm 87}}$ = $r$, and $A = rsA_1$.
\begin{figure}[tbh]
  \centering
   \includegraphics[width = 0.8\linewidth]{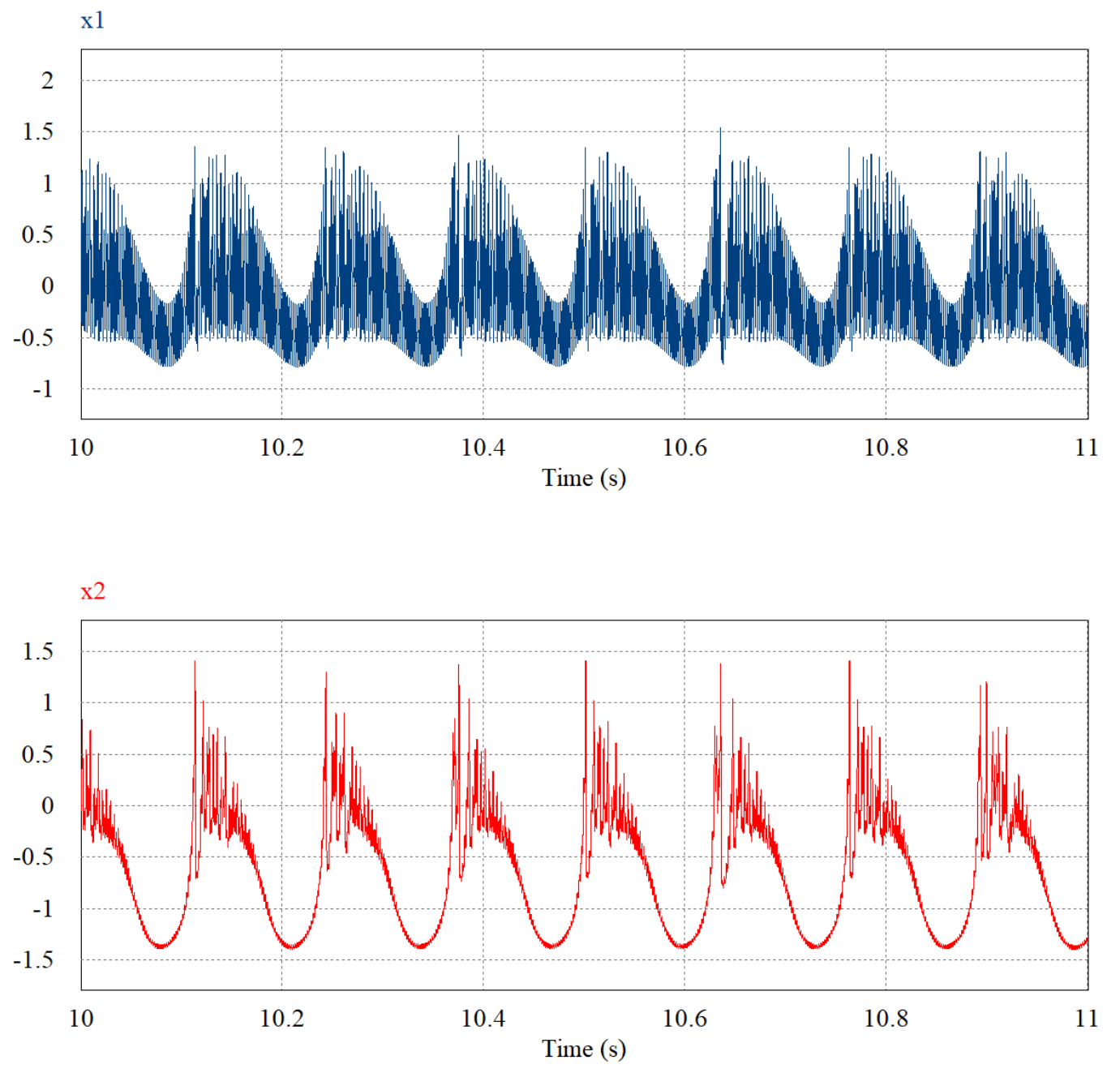}
  \caption{Time-Series waveforms of the system: (a) Time-Series of $x_1$ of the first oscillator. (b) Time-Series waveform of $x_2$ of the second oscillator. For each figure, $x$-axis is the time in seconds and the $y$-axis is the amplitudes of the state variables in V. The constant parameter values are $a_1$ = $a_2$ = $1$, $b_1$ = $b_2$ = $3$, $c_1$ = $c_2$ = $1$, $d_1$ = $d_2$ = $5$, $e = 0.5$, $r = 0.008$, $A = -1.6$, $s = 4$, $f = 500$~Hz. The varying parameter values are $m = 2$, $\sigma = 1$, $i = 3$, and $\alpha = 0.5$. The initial condition is chosen at ($-0.54$~V,$-5,0$~V, $0.1$~V,$0$~V,$0$~V,$6.688$~V). (Color online)}
 \label{fig:Fig 3}
\end{figure}

From Fig.~\ref{fig:Fig 1}, the outputs of the Op-Amps OA$2$, OA$9$, and OA$6$ are the state variables $x_1$, $y_1$, and $z_1$ of the first oscillator, respectively. Similarly, the outputs of the Op-Amps OA$11$, OA$14$, and OA$16$ are the state variables $x_2$, $y_2$, and $z_2$ of the second oscillator, respectively. The state variables from the circuit diagram have a unit, which is in volt. We have provided the initial conditions as initial voltages of the six capacitors $C_1$ to $C_6$. The ac sine wave in the expression \ref{eq3} is expressed here in terms of $V$. In case of non-dimensional equation~(\ref{eq3}), the value of frequency $f$ is $0.5$. But the circuit diagram equations are the dimensional one. So, to maintain the equivalence between them, the frequency of $V$ is chosen $500$~Hz. To simulate the system in the PSIM platform, we have chosen the time-step as $5~\mu$s, total time as $11$~sec. We have eliminated the first $10$~sec as the transient-time and have plotted the last second. We have used unity gain inverting amplifiers to achieve the state variables' negative values.

In order to obtain the experimental validations of the numerical predictions, we have varied $i$ and $\sigma$, keeping the remaining parameters fixed.

Fig.~\ref{fig:Fig 3} is the time-series waveforms of the two oscillators. Fig.~\ref{fig:Fig 3}(a) is the time-series waveform of the state variable $x_1$ of the first oscillator. Fig.~\ref{fig:Fig 3}(b) is the time-series waveform of the state variable $x_2$ of the second oscillator. The two figures are well-agreed with the numerical results as shown in Fig.~\ref{fig4}(c) and \ref{fig4}(d) in the parameter values as mentioned in the caption.

\begin{figure}[tbh]
\centering
\includegraphics[width = 0.8\linewidth]{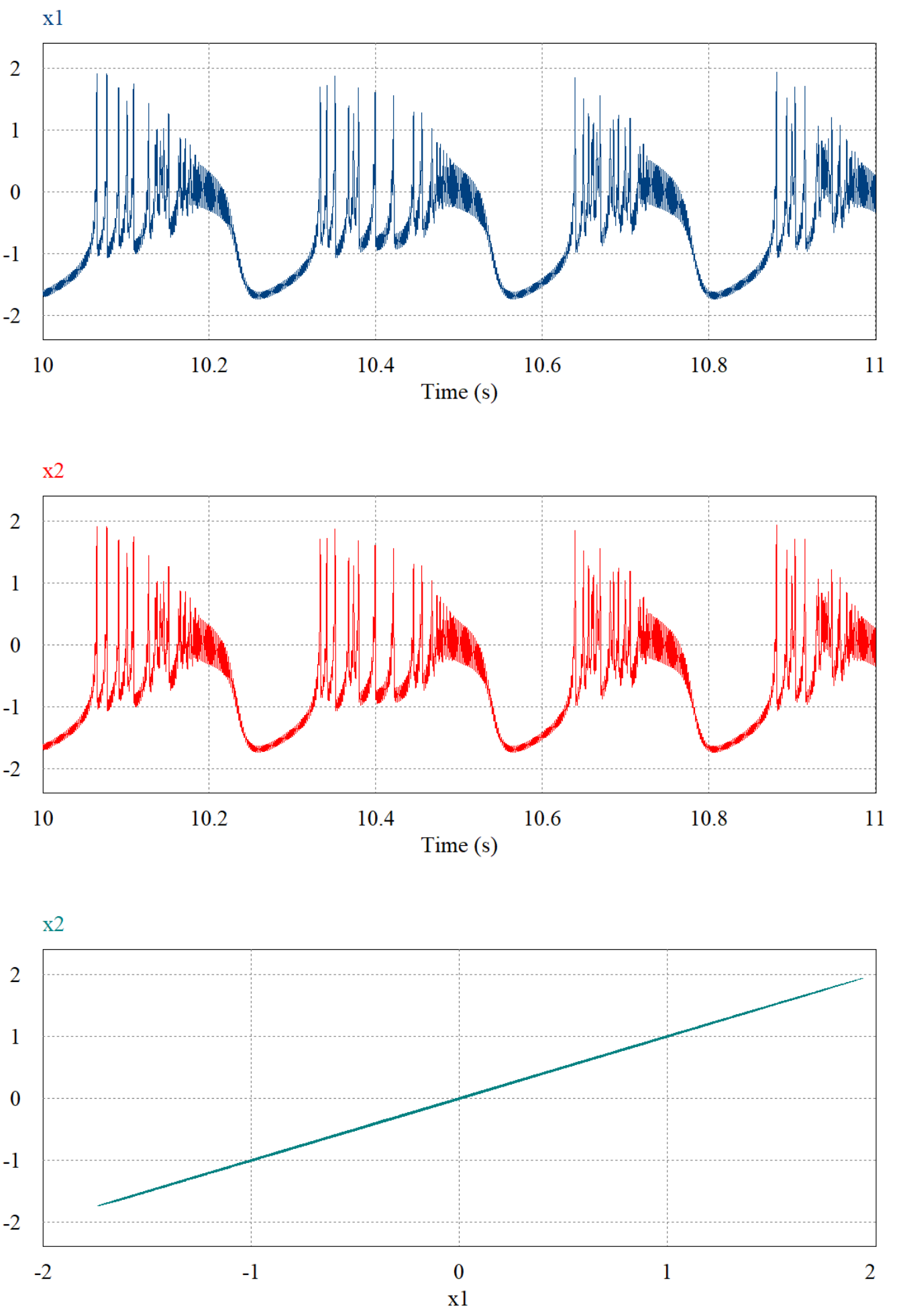}
\caption{Time-Series waveforms of the system: (a) Time-Series of $x_1$ the first oscillator. (b) Time-Series waveform of $x_2$ of the second oscillator. For each figure, $x$-axis is the time in seconds and the $y$-axis is the amplitudes of the state variables in V. The constant parameter values are $a_1$ = $a_2$ = $1$, $b_1$ = $b_2$ = $3$, $c_1$ = $c_2$ = $1$, $d_1$ = $d_2$ = $5$, $e = 0.5$, $r = 0.008$, $A = -1.6$, $s = 4$, $f = 500$~Hz. The varying parameter values are $m = 2$, $\sigma = 100$, $i = 3$, and $\alpha = 0.5$. The initial condition is chosen at ($-0.54$~V,$-5,0$~V, $0.1$~V,$0$~V,$0$~V,$6.688$~V). (Color online)}
\label{fig:Fig 3.1}
\end{figure}
Fig.~\ref{fig:Fig 3.1} is the time-series waveforms of the two oscillators. Fig.~\ref{fig:Fig 3.1}(a) is the time-series waveform of the state variable $x_1$ of the first oscillator. Fig.~\ref{fig:Fig 3.1}(b) is the time-series waveform of the state variable $x_2$ of the second oscillator. The two figures are well-agreed with the numerical results in the parameter values as mentioned in the caption. Looking at the two waveforms carefully, we find that the waveforms are identical, making the two oscillators synchronized in the parameter ranges mentioned in the caption. If we plot the phase-space, $x_1$-$x_2$, we confirm that the first state variable of the two oscillators, i.e., $x_1$ and $x_2$, are synchronized.

\begin{figure}[tbh]
\centering
\includegraphics[width = 0.8\linewidth]{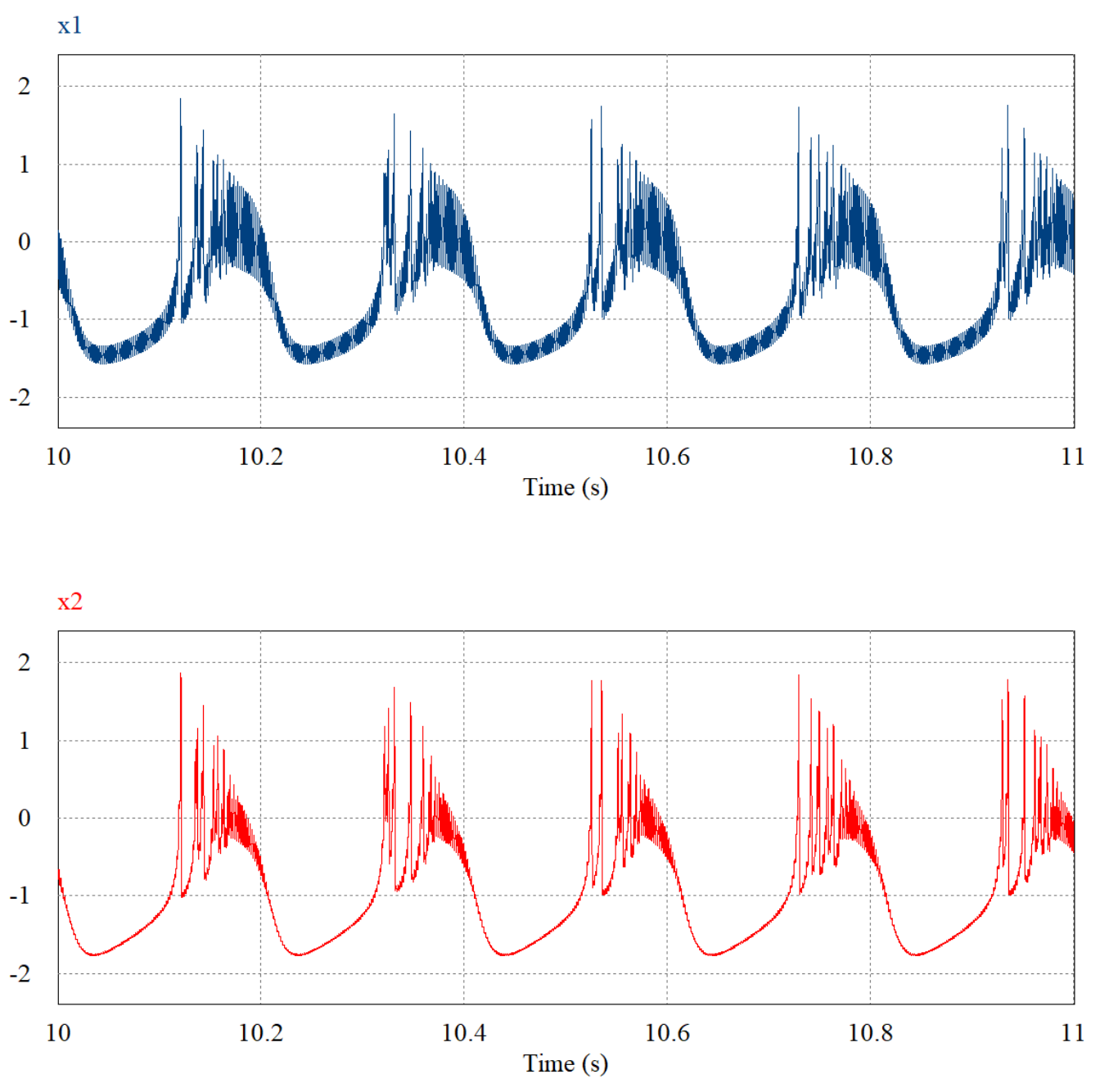}
\caption{Time-Series waveforms of the system: (a) Time-Series of $x_1$ the first oscillator. (b) Time-Series waveform of $x_2$ of the second oscillator. For each figure, $x$-axis is the time in seconds and the $y$-axis is the amplitudes of the state variables in V. The constant parameter values are $a_1$ = $a_2$ = $1$, $b_1$ = $b_2$ = $3$, $c_1$ = $c_2$ = $1$, $d_1$ = $d_2$ = $5$, $e = 0.5$, $r = 0.008$, $A = -1.6$, $s = 4$, $f = 500$~Hz. The varying parameter values are $m = 2$, $\sigma = 2$, $i = 2$, and $\alpha = 0.5$. The initial condition is chosen at ($-0.54$~V,$-5,0$~V, $0.1$~V,$0$~V,$0$~V,$6.688$~V). (Color online)}
 \label{fig:Fig 4}
\end{figure}
Fig.~\ref{fig:Fig 4} is the time-series waveforms of the two oscillators. Fig.~\ref{fig:Fig 4}(a) is the time-series waveform of the state variable $x_1$ of the first oscillator. Fig.~\ref{fig:Fig 4}(b) is the time-series waveform of the state variable $x_2$ of the second oscillator. The two figures are well-agreed with the numerical results in the parameter values as mentioned in the caption.

\section{Conclusion}
\label{conclusion}
This paper explores the collective behavior of two coupled and ring-star networks of heterogeneous neurons consisting of memristive $2$-D HR and traditional $3$-D HR neurons. The dynamical behavior of two heterogeneous coupled neurons via an electrical synapse has been investigated using the nonlinear analysis techniques based on the two-parameters maximal Lyapunov exponent graphs, bifurcation diagrams, and time-series waveforms. It has been found that the coupled model was able to exhibit a rich kind of dynamics, such as periodic, quasi-periodic, and chaotic dynamics involving bursting and spiking oscillations when parameters were varied smoothly in one direction. The absence of synchronization was observed for weak electrical coupling strength, while a synchronization cluster was observed between those neurons for higher values of the coupling strength. We have considered three different topological heterogeneous configurations concerning the collective behavior of up to $100$ coupled neurons in a ring-star network. We also investigated various spatiotemporal patterns exhibited by them. It was found that adding heterogeneity to the system increases the prevalence of chimera states and cluster states in the network. It also shows the prevalence of a new clustered chimera state. To further support the fact that our obtained result was not related to an artifact, the electronic circuit of the electronic heterogeneous coupled neurons was built and run in a PSIM environment to support our numerically obtained results further.

\begin{acknowledgements}
This work is partially funded by the Polish National Science Center under the Grant OPUS $ 14 No.2017/27/B/ST8/01330 $.
\end{acknowledgements}

%
 \section*{Compliance with ethical standards}

 \section*{Conflict of interest}
 The authors declare that they have no conflict of interest. 

\begin{thebibliography}{10}
\providecommand{\url}[1]{{#1}}
\providecommand{\urlprefix}{URL }
\expandafter\ifx\csname urlstyle\endcsname\relax
  \providecommand{\doi}[1]{DOI~\discretionary{}{}{}#1}\else
  \providecommand{\doi}{DOI~\discretionary{}{}{}\begingroup
  \urlstyle{rm}\Url}\fi

\bibitem{ref33}
Aghababaei, S., Balaraman, S., Rajagopal, K., Parastesh, F., Panahi, S.,
  Jafari, S.: Effects of autapse on the chimera state in a hindmarsh-rose
  neuronal network.
\newblock Chaos, Solitons \& Fractals  (2021)

\bibitem{ref1}
Angevine, J.: Encyclopedia of the human brain (2002)

\bibitem{ref2}
Bahramian, A., Parastesh, F., Pham, V.T., Kapitaniak, T., Jafari, S., Perc, M.:
  Collective behavior in a two-layer neuronal network with time-varying
  chemical connections that are controlled by a petri net.
\newblock Chaos: An Interdisciplinary Journal of Nonlinear Science
  \textbf{31}(3), 033138 (2021)

\bibitem{banerjee1998robust}
Banerjee, S., Yorke, J.A., Grebogi, C.: Robust chaos.
\newblock Physical Review Letters \textbf{80}(14), 3049 (1998)

\bibitem{ref27}
Bao, B., Yang, Q., Zhu, L., Bao, H., Xu, Q., Yu, Y., Chen, M.: Chaotic bursting
  dynamics and coexisting multistable firing patterns in 3d autonomous
  morris-lecar model and microcontroller-based validations.
\newblock Int. J. Bifurc. Chaos \textbf{29}, 1950134:1--1950134:18 (2019)

\bibitem{ref15}
Chay, T.R.: Chaos in a three-variable model of an excitable cell.
\newblock Physica D: Nonlinear Phenomena \textbf{16}, 233--242 (1985)

\bibitem{ref23}
Doubla, I.S., Ramakrishnan, B., Njitacke, Z.T., Kengne, J., Rajagopal, K.:
  Hidden extreme multistability and its control with selection of a desired
  attractor in a non-autonomous hopfield neuron.
\newblock AEU - International Journal of Electronics and Communications  (2021)

\bibitem{ref12}
Doubla, I.S., Ramakrishnan, B., Njitacke, Z.T., Kengne, J., Rajagopal, K.:
  Hidden extreme multistability and its control with selection of a desired
  attractor in a non-autonomous hopfield neuron.
\newblock AEU-International Journal of Electronics and Communications
  \textbf{144}, 154059 (2022)

\bibitem{ref3}
Gonzalez, J.: History-dependence of neuromodulation reduces levels of chaos in
  neuronal transitions.
\newblock Ph.D. thesis, Illinois State University (2021)

\bibitem{ref19}
Hindmarsh, J.L., Rose, R.M.: A model of the nerve impulse using two first-order
  differential equations.
\newblock Nature \textbf{296}, 162--164 (1982)

\bibitem{ref20}
Hindmarsh, J.L., Rose, R.M.: A model of neuronal bursting using three coupled
  first order differential equations.
\newblock Proceedings of the Royal Society of London. Series B. Biological
  Sciences \textbf{221}, 102 -- 87 (1984)

\bibitem{ref14}
Hodgkin, A.L., Huxley, A.F.: A quantitative description of membrane current and
  its application to conduction and excitation in nerve.
\newblock The Journal of Physiology \textbf{117} (1952)

\bibitem{ref31}
Hussain, I., Jafari, S., Ghosh, D., Perc, M.: Synchronization and chimeras in a
  network of photosensitive fitzhugh–nagumo neurons.
\newblock Nonlinear Dynamics  (2021)

\bibitem{ref30}
Hussain, I., Jafari, S., Perc, M., Ghosh, D.: Chimera states in a
  multi-weighted neuronal network.
\newblock Physics Letters A  (2021)

\bibitem{ref16}
Izhikevich, E.M.: Simple model of spiking neurons.
\newblock IEEE transactions on neural networks \textbf{14 6}, 1569--72 (2003)

\bibitem{ref8}
Li, Y.: Simulation of memristive synapses and neuromorphic computing on a
  quantum computer.
\newblock Physical Review Research \textbf{3}(2), 023146 (2021)

\bibitem{ref26}
Li, Z., Guo, Z., Wang, M., Ma, M.: Firing activities induced by memristive
  autapse in fitzhugh–nagumo neuron with time delay.
\newblock AEU - International Journal of Electronics and Communications  (2021)

\bibitem{Mu20a}
Muni, S.S., Provata, A.: Chimera states in ring–star network of chua
  circuits.
\newblock Nonlinear Dyn \textbf{101}, 2509--2521 (2020)

\bibitem{ref29}
Muni, S.S., Rajagopal, K., Karthikeyan, A., Arun, S.: Discrete hybrid
  izhikevich neuron model: Nodal and network behaviours considering
  electromagnetic flux coupling.
\newblock Chaos, Solitons \& Fractals  (2022)

\bibitem{ref13}
Njitacke, Z.T., Awrejcewicz, J., Ramakrishnan, B., Rajagopal, K., Kengne, J.:
  Hamiltonian energy computation and complex behavior of a small heterogeneous
  network of three neurons: circuit implementation.
\newblock Nonlinear Dynamics pp. 1--20 (2021)

\bibitem{ref11}
Njitacke, Z.T., Isaac, S.D., Nestor, T., Kengne, J.: Window of multistability
  and its control in a simple 3d hopfield neural network: application to
  biomedical image encryption.
\newblock Neural Computing and Applications \textbf{33}(12), 6733--6752 (2021)

\bibitem{ref10}
Njitacke, Z.T., Tsafack, N., Ramakrishnan, B., Rajagopal, K., Kengne, J.,
  Awrejcewicz, J.: Complex dynamics from heterogeneous coupling and
  electromagnetic effect on two neurons: Application in images encryption.
\newblock Chaos, Solitons \& Fractals \textbf{153}, 111577 (2021)

\bibitem{ref36}
Potapov, A., Ali, M.K.: Robust chaos in neural networks.
\newblock Physics Letters A \textbf{277}, 310--322 (2000)

\bibitem{ref17}
Quininao, C., Touboul, J.D.: Clamping and synchronization in the strongly
  coupled fitzhugh--nagumo model.
\newblock SIAM Journal on Applied Dynamical Systems \textbf{19}(2), 788--827
  (2020)

\bibitem{seth2019observation}
Seth, S.: Observation of robust chaos in 3d electronic system.
\newblock IET Circuits, Devices \& Systems \textbf{13}(4), 558--564 (2019)

\bibitem{Sh17a}
Shepelev, I., Vadivasova, T., Bukh, A., Strelkova, G., Anishchenko, V.: New
  type of chimera structures in a ring of bistable fitzhugh–nagumo
  oscillators with nonlocal interaction.
\newblock Physics Letters A \textbf{381}(16), 1398--1404 (2017)

\bibitem{ref34}
Simo, G.R., Louodop, P., Ghosh, D., Njougouo, T., Tchitnga, R., Cerdeira, H.A.:
  Traveling chimera patterns in a two-dimensional neuronal network.
\newblock p. 127519. Elsevier (2021)

\bibitem{ref35}
Simo, G.R., Njougouo, T., Aristides, R., Louodop, P., Tchitnga, R., Cerdeira,
  H.A.: Chimera states in a neuronal network under the action of an electric
  field.
\newblock Physical Review E \textbf{103}(6), 062304 (2021)

\bibitem{ref24}
Telksnys, T., Navickas, Z., Timofejeva, I., Marcinkevicius, R., Ragulskis, M.:
  Symmetry breaking in solitary solutions to the hodgkin–huxley model.
\newblock Nonlinear Dynamics pp. 1--12 (2019)

\bibitem{ref18}
Tsumoto, K., Kitajima, H., Yoshinaga, T., Aihara, K., Kawakami, H.:
  Bifurcations in morris-lecar neuron model.
\newblock Neurocomputing \textbf{69}, 293--316 (2006)

\bibitem{ref5}
Wang, W., He, C., Wang, Z., Cheng, J., Mo, X., Tian, K., Fan, D., Luo, X.,
  Yuan, M., Kurths, J.: Dynamic analysis of disease progression in
  alzheimer’s disease under the influence of hybrid synapse and spatially
  correlated noise.
\newblock Neurocomputing \textbf{456}, 23--35 (2021)

\bibitem{ref32}
Wang, Z., Xu, Y., Li, Y., Kapitaniak, T., Kurths, J.: Chimera states in coupled
  hindmarsh-rose neurons with $\alpha$-stable noise.
\newblock Chaos, Solitons \& Fractals \textbf{148}, 110976 (2021)

\bibitem{ref28}
Xu, L., Qi, G., Ma, J.: Modeling of memristor-based hindmarsh-rose neuron and
  its dynamical analyses using energy method.
\newblock Applied Mathematical Modelling  (2022)

\bibitem{ref22}
Xu, Q., Ju, Z., Ding, S., Feng, C., Chen, M., Bao, B.: Electromagnetic
  induction effects on electrical activity within a memristive wilson neuron
  model.
\newblock Cognitive Neurodynamics pp. 1--11 (2022)

\bibitem{ref21}
Xu, Q., Liu, T., Feng, C.T., Bao, H., Wu, H.G., Bao, B.C.: Continuous
  non-autonomous memristive rulkov model with extreme multistability.
\newblock Chinese Physics B \textbf{30}(12), 128702 (2021)

\bibitem{ref25}
Xu, Q., Tan, X., Zhu, D., Bao, H., Hu, Y., Bao, B.: Bifurcations to bursting
  and spiking in the chay neuron and their validation in a digital circuit.
\newblock Chaos Solitons \& Fractals \textbf{141}, 110353 (2020)

\bibitem{ref6}
Yao, Z., Zhou, P., Zhu, Z., Ma, J.: Phase synchronization between a
  light-dependent neuron and a thermosensitive neuron.
\newblock Neurocomputing \textbf{423}, 518--534 (2021)

\bibitem{ref9}
Zhang, G., Guo, D., Wu, F., Ma, J.: Memristive autapse involving magnetic
  coupling and excitatory autapse enhance firing.
\newblock Neurocomputing \textbf{379}, 296--304 (2020)

\bibitem{ref7}
Zhang, Y., Wang, C., Tang, J., Ma, J., Ren, G.: Phase coupling synchronization
  of fhn neurons connected by a josephson junction.
\newblock Science China Technological Sciences \textbf{63}(11), 2328--2338
  (2020)

\bibitem{ref4}
Zhou, J.F., Jiang, E.H., Xu, B.L., Xu, K., Zhou, C., Yuan, W.J.: Synaptic
  changes modulate spontaneous transitions between tonic and bursting neural
  activities in coupled hindmarsh-rose neurons.
\newblock Physical Review E \textbf{104}(5), 054407 (2021)

\end{thebibliography}

\end{document}